%% file: main.tex
\documentclass[aps,pra,reprint,longbibliography,floatfix]{revtex4-2}

\usepackage[T1]{fontenc}
\usepackage[utf8]{inputenc}
\usepackage{lmodern}
\usepackage{microtype}
\usepackage{amsmath, amssymb}
\usepackage{graphicx}
\usepackage{tikz}
\usetikzlibrary{arrows.meta,calc}
\usepackage{booktabs}
\usepackage{xcolor}
\usepackage[colorlinks=true, linkcolor=blue, citecolor=blue, urlcolor=blue]{hyperref}

\graphicspath{{figures/}}

\newcommand{\Ssys}{\mathcal{S}}
\newcommand{\Env}{\mathcal{E}}
\newcommand{\Frag}{\mathcal{F}}
\newcommand{\Rem}{\mathcal{R}}
\newcommand{\ket}[1]{\lvert #1 \rangle}
\newcommand{\bra}[1]{\langle #1 \rvert}

\begin{document}

\title{Environment Alignment and Redundant Record Formation in Imperfect-CNOT Quantum Darwinism}
\author{Aleksander Lasek}
\affiliation{International Centre for Theory of Quantum Technologies, University of Gdańsk, 80-309 Gdańsk, Poland}
\author{Paweł Horodecki}
\affiliation{International Centre for Theory of Quantum Technologies, University of Gdańsk, 80-309 Gdańsk, Poland}
\date{September 14, 2026}

\input{sections/abstract}

\maketitle

\input{sections/introduction}

\input{sections/methods}

\input{sections/results}

\input{sections/discussion}

\input{sections/conclusions}

\input{sections/acknowledgments}

\input{sections/data_availability}

\bibliography{paper_a_bibliography}

\input{sections/supplement}

\end{document}

%% file: sections/abstract.tex
\begin{abstract}
Quantum Darwinism explains objective information through redundant
environmental records. Earlier work established that imperfect records can
be amplified and that environment self-evolution can enhance or suppress
their formation. We investigate how preparation, interaction angle, and
field disorder combine in an imperfect-CNOT model with random Gaussian couplings
and pure, noninteracting environment qubits. We find that, without fields,
a single alignment parameter $\Lambda$ determines the preparation and
interaction-angle dependence of conditional-state distinguishability.
Increasing interaction imperfection or local field strength
can improve or suppress recording. We explain this nonmonotonic response
through the geometry of conditional branch separation. For the pure initial
environments considered here, the $Z$ basis remains optimal for Holevo
information, so field-assisted recording requires no change of the recorded
system observable. Comparing uniform local field strengths with
Gaussian-distributed strengths at equal root-mean-square strength shows how disorder
broadens both beneficial and detrimental field effects. Exact expressions
for fragment information and numerical simulations with up to 24 environment qubits
quantify the distinction between high mean information and reliable records
across fragments and throughout a finite observation window. These results
connect the geometry of local information acquisition to the fragment sizes
needed for robust recording. The branch-distinguishability analysis itself
requires only the conditional evolution of each environment qubit and extends
to other interactions that preserve a system pointer basis.
\end{abstract}

%% file: sections/introduction.tex
\section{Introduction}
\label{sec:introduction}

Quantum Darwinism (QD) provides an operational account of how objective classical information can emerge from quantum dynamics. Rather than interacting directly with a system $\Ssys$, observers usually intercept fragments $\Frag$ of its environment $\Env$. A property of $\Ssys$ becomes objective when many observers can independently infer the same value from disjoint fragments, without appreciably perturbing either the system or one another's records \cite{OllivierPoulinZurek2004ObjectiveProperties,OllivierPoulinZurek2005EnvironmentWitness,Zurek2009QuantumDarwinism}. Such disjoint readouts can be viewed as compatible indirect measurements of the system \cite{DoucetDeffner2025Compatibility}. The environment then acts not only as a sink for coherence, but also as a communication channel that selectively proliferates information about a stable system observable \cite{Zurek2003Decoherence}.
These signatures have been observed experimentally, including in a six-photon
quantum simulator \cite{ChenZhongLiEtAl2019PhotonicSimulator}, on a
superconducting processor \cite{ZhuSaliceTouilEtAl2025Superconducting}, and
on trapped-ion and superconducting devices
\cite{DoucetDeffner2026NISQ}.  The superconducting-circuit experiment of
Ref.~\cite{ZhuSaliceTouilEtAl2025Superconducting} also separates the fragment
mutual information into classically accessible and discord contributions.

The standard partial-information diagnostic is the quantum mutual information between the system and a fragment,
\begin{equation}
    I(\Ssys:\Frag)=S(\rho_{\Ssys})+S(\rho_{\Frag})-S(\rho_{\Ssys\Frag}),
    \label{eq:intro_mutual_information}
\end{equation}
where $S(\rho)=-\operatorname{Tr}(\rho\log_2\rho)$ is the von Neumann
entropy. In a Darwinistic branching state, $I(\Ssys:\Frag)$ rises rapidly to a plateau
near the pointer-value entropy of the decohered system, $H_Z(\Ssys)$: a small
fragment already supplies nearly all of the classical information about the
system \cite{BlumeKohoutZurek2006QuantumDarwinism,TouilYanGirolamiEtAl2022Eavesdropping}.

Conventionally, one defines $m_\delta$ as the smallest fragment size for which
\begin{equation}
    \overline{I(\Ssys:\Frag_{m_\delta})}\geq(1-\delta)H_Z(\Ssys),
    \qquad
    R_\delta=\frac{N_{\Env}}{m_\delta},
    \label{eq:intro_redundancy}
\end{equation}
where $N_{\Env}$ is the number of environment subsystems and the overline denotes an average over fragments of the stated size \cite{BlumeKohoutZurek2006QuantumDarwinism,TouilYanGirolamiEtAl2022Eavesdropping}. Large $R_\delta$ estimates record multiplicity under the stated
fragment statistic; by itself it does not construct a disjoint partition
whose blocks all pass the threshold. In this work we apply the analogous threshold to the pointer-basis Holevo information introduced below, because the mutual information alone does not separate classical records from other correlations.

The observable whose values are redundantly recorded is not arbitrary.
Decoherence selects stable pointer states through the system--environment
interaction \cite{Zurek1981PointerBasis}; in the model studied here, the
system enters the Hamiltonian only through a projector onto one of its two
states, and that projector fixes the pointer observable
\cite{DuruisseauTouilDeffner2023PointerStates,DoucetDeffner2024Classifying}.
That observable is $\sigma_z^{\Ssys}$, with pointer states
$\{|0\rangle_{\Ssys},|1\rangle_{\Ssys}\}$.  Changing the conditional transformation of
an environment qubit can change how well it distinguishes these two pointer
states without changing the pointer states themselves.

Mutual information alone does not distinguish classical pointer records from
nonclassical correlations, and a Darwinistic plateau need not imply spectrum
broadcast structure (SBS) or operational objectivity
\cite{KorbiczHorodeckiHorodecki2014Broadcasting,HorodeckiKorbiczHorodecki2015QuantumOrigins,LeOlayaCastro2018Objectivity,LeOlayaCastro2019StrongQD,Korbicz2021Roads}.
In an SBS, multiple disjoint environment fragments
carry mutually distinguishable records of the same classical system value and
are conditionally independent once that value is specified.  We therefore
separate the pointer-basis Holevo information from the remaining correlations:
\begin{align}
    \chi_z(\Ssys:\Frag)
    &=S(\rho_{\Frag})-\sum_{s=0}^{1}p_sS(\rho_{\Frag|s}),
    \\
    D_z(\Ssys:\Frag)
    &=I(\Ssys:\Frag)-\chi_z(\Ssys:\Frag).
    \label{eq:intro_holevo_discord}
\end{align}
Here $H_Z(\Ssys)=-\sum_s p_s\log_2p_s$ denotes the Shannon entropy of the
system's pointer-value distribution; it is one bit for the equal branch weights
used below.  The quantity $\chi_z$ is the Holevo information of the conditional
fragment ensemble and therefore upper-bounds the information recoverable by
measuring $\Frag$, while $D_z$ is the remainder left when the system
measurement is fixed to the pointer basis.  It is discord-like, but it is not
the discord minimized over all measurements
\cite{HendersonVedral2001Correlations,OllivierZurek2001Discord,ZwolakZurek2013DiscordAccessible}.
We use the Holevo information to quantify the pointer information
carried by environment fragments. Redundant recording requires this
information to approach $H_Z(\Ssys)$ already for small fragments and remain
near saturation as fragment size increases. A small pointer-basis remainder
over this range gives the corresponding classical plateau in mutual
information. Small $D_z$ alone is inconclusive: it can describe either a clean
classical record or a fragment containing almost no information.

Imperfect environmental records can be amplified into redundant
information, as demonstrated explicitly for the imperfect-CNOT (C-MAYBE) gate by Touil
\emph{et al.}~\cite{TouilYanGirolamiEtAl2022Eavesdropping}.
Zwolak, Riedel, and Zurek analyzed spin environments with nonidentical
couplings and local self-evolution, including preparation directions that
acquire no information and fields that enhance or suppress recording
\cite{ZwolakRiedelZurek2016SpinEnvironments}.
Mironowicz \emph{et al.} in their few-qubit analysis combined the same gate, which they call the
controlled imperfect-NOT (C-INOT), with environment self-evolution, mixed
initial environments, and selected interactions within the environment
\cite{MironowiczHorodeckiHorodecki2022SelfEvolution}. Their
calculations of the distance to the nearest SBS state showed a nonmonotonic
dependence on both self-evolution and mixedness, and motivated a generalized
pointer, or indicator, basis.

Here we connect these perspectives through a systematic analysis
of preparation, interaction angle, and local fields in a pure,
noninteracting environment. Our contribution is to resolve these dependences
for the C-INOT family and compare field distributions and fragment
reliability within the same model. We report three main results.
\begin{enumerate}
    \item Preparation and interaction angle admit a common
    description. Without fields, their effect on the conditional-state
    overlap is captured by the alignment parameter $\Lambda$, the squared
    sine of the angle between the initial environment Bloch vector and the
    conditional rotation axis. Its zero value identifies preparations that
    write no record. This makes the known role of preparation alignment
    \cite{ZwolakQuanZurek2010RedundantImprinting} explicit across the gate
    family and relates apparently different parameter choices.

    \item Recording responds nonmonotonically to interaction
    imperfection and local field strength: increasing either can improve or
    suppress it. Relative conditional rotations explain this response through
    changes in the distinguishability of the environment states conditioned
    on the two system pointer values.
    We compare a common
    field strength on all environment qubits with local strengths sampled
    independently from a zero-mean Gaussian distribution. Matching the
    root-mean-square strength separates the effects of field strength and
    field disorder.

    \item We find that high mean information can conceal fragments
    carrying little information and temporary losses of recording.
    Requiring most fragments to meet the information threshold increases
    the necessary fragment size. Using the product of local overlaps
    underlying established amplification results
    \cite{ZwolakRiedelZurek2014Chernoff,TouilYanGirolamiEtAl2022Eavesdropping},
    we quantify this increase. For suitable fields, records remain
    available across most half-environment fragments at every sampled time in a late
    observation window, selected after the initial no-field dephasing
    transient.
\end{enumerate}

Numerically, we draw the system--environment couplings at random and hold them
fixed during each evolution, for two reasons.  Without self-fields,
identically prepared environment qubits, or witnesses, with identical couplings oscillate in phase
and produce periodic fragment-information revivals.  Random couplings
dephase the no-field disorder averages toward a stationary late-time limit.
A finite closed realization still exhibits quasiperiodic dynamics.  They also make record quality vary among environment qubits,
allowing us to identify fragments carrying little information even
when the mean is high.
The conclusions are based on independent coupling realizations and checked
against exact averages over the coupling distribution, rather than inferred
from a selected realization.

The main claims concern pure product initial environments with no
environment--environment interactions, $H_{\Env\Env}=0$.  The
system--environment couplings are nonuniform, and some calculations add either
uniform or random local fields to the environment qubits; comparing the two
separates the effect of field strength from that of field inhomogeneity.
Pure states isolate
the alignment mechanism from the loss of record capacity caused by initial
environment mixedness \cite{ZwolakQuanZurek2009MixedEnvironment}.
We also derive and numerically check how mixedness changes branch
separation and information in individual witnesses
(Sec.~\ref{subsec:discussion_mironowicz} and
Fig.~\ref{fig:supp_mixed_comparison}). Larger-fragment redundancy and
persistence for mixed environments, and the effects of environment
interactions, remain for future work.

%% file: sections/methods.tex
\section{Methods}
\label{sec:methods}

\subsection{Imperfect record writing}
\label{subsec:methods_mironowicz}
We study record formation in a controlled-unitary model: the system's two pointer
values condition two different transformations of each environment qubit.  A
qubit records the pointer value only insofar as these transformations produce
distinguishable conditional states.  This setting makes it possible to
separate the nonlocal content of the interaction from the ability of a
particular environment preparation to register that interaction. 
More specifically, we use the $\theta$-parametrized CNOT
system--environment coupling analyzed first in the context of quantum Darwinism
in Ref.~\cite{TouilYanGirolamiEtAl2022Eavesdropping} and later also applied
to small mixed-state environments in
Ref.~\cite{MironowiczHorodeckiHorodecki2022SelfEvolution}.
Namely, the controlled gate in this model can be written
\begin{equation}
    U_\theta
    =|0\rangle\!\langle0|\otimes I
    +|1\rangle\!\langle1|\otimes
      (\cos\theta\,\sigma_x+\sin\theta\,\sigma_z).
    \label{eq:methods_cinot}
\end{equation}
At $\theta=0$ this is CNOT, whereas at $\theta=\pi/2$ it is controlled-$Z$, not a maximally degraded CNOT. Although $\theta$ is conventionally called an imperfection parameter, it does not continuously weaken the nonlocal gate. With $V_\theta=e^{+i\theta\sigma_y/2}$,
\begin{equation}
    U_\theta=(I\otimes V_\theta)U_{\mathrm{CNOT}}(I\otimes V_\theta^\dagger),
    \label{eq:methods_local_equivalence}
\end{equation}
so the full family is locally equivalent to CNOT in the standard classification of two-qubit gates \cite{Makhlin2002NonlocalProperties,ZhangValaSastryWhaley2003Geometric}. The physical situations are nevertheless distinct, because the rotation $V_\theta$ is not applied to the environment's initial state, self-Hamiltonian, or measurement basis; only the axis of the conditional rotation changes.

Touil \emph{et al.} call this imperfect-CNOT family C-MAYBE
\cite{TouilYanGirolamiEtAl2022Eavesdropping}, whereas Mironowicz, Horodecki,
and Horodecki use controlled imperfect-NOT (C-INOT)
\cite{MironowiczHorodeckiHorodecki2022SelfEvolution}.  We use C-INOT below.
The nonmonotonic self-evolution response found in the latter work
(Sec.~\ref{sec:introduction}) raises the question addressed here:
\textit{does the nonmonotonicity indicate
that the environment records a different system observable, or does
self-evolution merely change the distinguishability of the environment
states conditioned on the two system pointer values?}  We use branch
separation to describe how distinguishable these conditional states become,
specifying whether we consider one witness, a fragment, or the full
environment. We test the second explanation for pure product
environment preparations.  For a fixed direction of an environment
qubit's initial Bloch vector, reducing its purity cannot increase the
information recorded about the system's $Z$ pointer value
\cite{ZwolakQuanZurek2009MixedEnvironment}, so this
restriction isolates the alignment mechanism.  Our main claims therefore
concern pure, initially uncorrelated environment qubits; their extension to
mixed environment qubits is discussed in
Sec.~\ref{subsec:discussion_mironowicz}.  We combine exact single-qubit
calculations with numerical simulations at nonuniform
system--environment couplings.  We focus on how the records are written; the
amplification of imperfect local records is already well understood
\cite{ZwolakQuanZurek2010RedundantImprinting,RiedelZurek2010Everyday}.

\subsection{System, environment, and Hamiltonian}
\label{subsec:methods_model}

We consider one system qubit $\Ssys$ coupled in a star geometry to
$N_{\Env}$ environment qubits $\Env_j$, which we also call witnesses
[Fig.~\ref{fig:methods_model}(a)].  Their joint Hilbert space is
\begin{align}
    \mathcal H
    &=\mathcal H_{\Ssys}\otimes\bigotimes_{j=1}^{N_{\Env}}\mathcal H_{\Env_j},
    \\
    \dim\mathcal H_{\Ssys}&=\dim\mathcal H_{\Env_j}=2,
    \qquad N_{\rm tot}=N_{\Env}+1.
    \label{eq:methods_hilbert_space}
\end{align}
We set $\hbar=1$ and write $P_1^{\Ssys}=|1\rangle\!\langle1|_{\Ssys}$. The generator of the conditional gate acting on environment qubit $j$ is
\begin{equation}
    H_\theta^{(j)}
    =P_1^{\Ssys}\otimes\frac{\pi}{2}
    \left(I^{(j)}-\cos\theta\,\sigma_x^{(j)}
    -\sin\theta\,\sigma_z^{(j)}\right),
    \label{eq:methods_pair_hamiltonian}
\end{equation}
so that unit coupling and unit evolution time give the gate $U_\theta$ in Eq.~\eqref{eq:methods_cinot}.

The model contains neither a system self-Hamiltonian nor interactions
within the environment.  Its full Hamiltonian is
\begin{align}
    H=\sum_{j=1}^{N_{\Env}}\Bigl[&
    g_j P_1^{\Ssys}\otimes\frac{\pi}{2}
    \bigl(I^{(j)}-\cos\theta\,\sigma_x^{(j)}
    \notag\\
    &\hspace{7.2em}-\sin\theta\,\sigma_z^{(j)}\bigr)
    +h_j I_{\Ssys}\otimes\sigma_z^{(j)}\Bigr],
    \label{eq:methods_hamiltonian}
\end{align}
where identity operators on environment qubits other than $j$ are implicit.
A system Hamiltonian is omitted: one diagonal in the pointer basis would
add only relative branch phases, while one that does not commute with
$\sigma_z^{\Ssys}$ would rotate the pointer states during record formation,
which is a different problem.

\input{sections/figure_model_schematic}

In the no-field model with identical couplings and identical witness
preparations, all witnesses oscillate in phase and the fragment information
shows periodic revivals.  Independent Gaussian couplings instead dephase the
corresponding disorder-averaged quantities and give the witnesses different
record qualities.  We
parameterize the couplings and local fields as
\begin{align}
    g_j&\overset{\mathrm{iid}}{\sim}\mathcal N(0,\sigma_g^2),
    & \sigma_g&=0.1,
    \\
    h_j&=h_0+W\xi_j,
    & \xi_j&\overset{\mathrm{iid}}{\sim}\mathcal N(0,1).
    \label{eq:methods_coupling_distributions}
\end{align}
All random variables $\{g_j,\xi_j\}$ are statistically independent, both
across environment qubits and between couplings and fields, so $W$ is the
standard deviation of the random local field.  The choices
$W=0$ and $h_0=0$ define, respectively, a uniform field and a zero-mean
Gaussian field; $W=h_0=0$ is the no-field model.  We express the Hamiltonian in an arbitrary energy unit $E_0$ and time
in units of $\hbar/E_0$.  Thus the plotted time is
$t=E_0t_{\rm phys}/\hbar$; converting it to seconds requires a physical
choice of $E_0$.  In the absence of self-fields, a coupling of magnitude
$|g_j|$ implements one C-INOT gate at $|g_j|t=1$.  The chosen width
$\sigma_g=0.1$ therefore defines a reference gate time of ten time units.
The intervals ending at $t=40$ and $60$ span four and six reference gate
times, respectively.

\subsection{Initial states and Bloch-vector parameterization}
\label{subsec:methods_initial_states}

The controlled Hamiltonian, rather than the initial state, selects the system
pointer observable $\sigma_z^{\Ssys}$.  We initialize the system in an equal
superposition of its pointer states,
\begin{equation}
    |\psi_{\Ssys}(0)\rangle=|+x\rangle
    =\frac{|0\rangle+|1\rangle}{\sqrt2},
    \label{eq:methods_initial_system_state}
\end{equation}
so that the two conditional branches have equal prior weights.  This choice
maximizes the available pointer information at one bit and prevents changes in
record quality from being obscured by a trivial imbalance between the two
system alternatives.  It also yields symmetric closed-form expressions for the
information diagnostics used below.

To study the geometry of the environment preparation without the separate loss
of information capacity caused by mixedness, we replace the diagonal mixed
environment state of Ref.~\cite{MironowiczHorodeckiHorodecki2022SelfEvolution}
with the pure product state
\begin{equation}
    |\Env(0)\rangle
    =\bigotimes_{j=1}^{N_{\Env}}|\phi(p)\rangle_j,
    \qquad
    |\phi(p)\rangle
    =\sqrt p\,|0\rangle+\sqrt{1-p}\,|1\rangle,
    \label{eq:methods_initial_environment_state}
\end{equation}
where $p$ is the $|0\rangle$ population. The corresponding Bloch vector is
\begin{equation}
    \mathbf r(p)
    =\left(2\sqrt{p(1-p)},0,2p-1\right).
    \label{eq:methods_initial_bloch_vector}
\end{equation}
Equivalently, setting $p=\cos^2(\beta/2)$ gives
\begin{equation}
    |\phi(\beta)\rangle
    =\cos\frac{\beta}{2}|0\rangle
    +\sin\frac{\beta}{2}|1\rangle,
    \qquad
    \mathbf r(\beta)=(\sin\beta,0,\cos\beta).
    \label{eq:methods_angular_environment_state}
\end{equation}

\subsection{Conditional branch geometry and information diagnostics}
\label{subsec:methods_geometry}

For an initial equal superposition of the system pointer states, the controlled
dynamics has the branching form
\begin{equation}
    |\Psi(t)\rangle=\frac{1}{\sqrt2}
    \left(|0\rangle_{\Ssys}|\Env_0(t)\rangle+|1\rangle_{\Ssys}|\Env_1(t)\rangle\right).
    \label{eq:methods_branching_state}
\end{equation}
Because $H$ is block diagonal in the system pointer basis, it decomposes into
two branch Hamiltonians acting on the environment alone,
\begin{equation}
    H=\sum_{s\in\{0,1\}}|s\rangle\!\langle s|_{\Ssys}\otimes H^{(s)},
    \qquad
    H^{(s)}=\sum_{j=1}^{N_{\Env}}H_j^{(s)},
    \label{eq:methods_branch_decomposition}
\end{equation}
with single-witness generators
\begin{equation}
\begin{aligned}
    H_j^{(0)}&=h_j\sigma_z^{(j)},
    \\
    H_j^{(1)}&=h_j\sigma_z^{(j)}
    +\frac{\pi g_j}{2}\left(I^{(j)}-\cos\theta\,\sigma_x^{(j)}
    -\sin\theta\,\sigma_z^{(j)}\right).
\end{aligned}
    \label{eq:methods_branch_generators}
\end{equation}
Hence $|\Env_s(t)\rangle=e^{-iH^{(s)}t}|\Env(0)\rangle$.  Because the
environment qubits do not interact, generators acting on different qubits
commute, and each branch propagator factorizes into single-qubit unitaries,
\begin{equation}
    e^{-iH^{(s)}t}=\bigotimes_{j=1}^{N_{\Env}}U_j^{(s)}(t),
    \qquad
    U_j^{(s)}(t)=e^{-iH_j^{(s)}t}.
    \label{eq:methods_branch_unitaries}
\end{equation}
Acting on the product initial state of
Eq.~\eqref{eq:methods_initial_environment_state}, each conditional environment
state remains a product,
\begin{equation}
    |\Env_s(t)\rangle
    =\bigotimes_{j=1}^{N_{\Env}}|e_{s,j}(t)\rangle,
    \qquad
    |e_{s,j}(t)\rangle=U_j^{(s)}(t)|\phi_j\rangle,
    \label{eq:methods_conditional_factorization}
\end{equation}
and the overlap of the two branches is a product of single-witness overlaps,
\begin{equation}
    \langle\Env_0(t)|\Env_1(t)\rangle
    =\prod_{j=1}^{N_{\Env}}
    \langle\phi_j|U_j^{(0)\dagger}(t)U_j^{(1)}(t)|\phi_j\rangle.
    \label{eq:methods_branch_overlap_product}
\end{equation}
Each witness thus enters only through its relative branch unitary
$U_j^{(0)\dagger}U_j^{(1)}$, and we define the local branch overlap
$B_j(t)=|\langle\phi_j|U_j^{(0)\dagger}(t)U_j^{(1)}(t)|\phi_j\rangle|$.
For pure initial environment states, $B_j$ is the root fidelity
between the two conditional states and also the modulus of the corresponding
local decoherence factor. These quantities underlie earlier analyses of
information acquisition and spectrum broadcasting
\cite{ZwolakRiedelZurek2016SpinEnvironments,MironowiczKorbiczHorodecki2017Monitoring}.
Here we evaluate them for the preparation and field configurations considered
below.

For environment qubit $j$, after stripping an irrelevant branch-dependent scalar phase, we write the $\mathrm{SU}(2)$ part of the relative branch evolution as
\begin{equation}
    U_j^{(0)\dagger}(t)U_j^{(1)}(t)
    =q_{0j}(t)I-i\mathbf q_j(t)\cdot\boldsymbol\sigma.
    \label{eq:methods_relative_unitary}
\end{equation}
Every $\mathrm{SU}(2)$ matrix has this axis--angle, or unit-quaternion, form
with real coefficients satisfying $q_{0j}^2+|\mathbf q_j|^2=1$.
Specifically, $q_{0j}=\cos(\varphi_j/2)$ and
$\mathbf q_j=\hat{\mathbf u}_j\sin(\varphi_j/2)$ for a relative Bloch-sphere
rotation through angle $\varphi_j$ about axis $\hat{\mathbf u}_j$.  When
$\mathbf q_j\neq0$, its direction gives the relative rotation axis; its
magnitude is $\sin(\varphi_j/2)$.
If its initial pure state has Bloch vector $\mathbf r_j$, the conditional branch overlap is
\begin{equation}
\begin{aligned}
    B_j^2(t)
    &=q_{0j}^2(t)+[\mathbf q_j(t)\cdot\mathbf r_j]^2,
    \\
    &=1-A_j(t),
    \\
    A_j(t)&=|\mathbf q_j(t)\times\mathbf r_j|^2.
\end{aligned}
    \label{eq:methods_record_separation}
\end{equation}
We quantify local branch separation by $A_j=1-B_j^2$: it is zero
for identical conditional states of witness $j$ and one for orthogonal states.
The vector $\mathbf q_j$ characterizes the relative conditional rotation,
while $A_j$ is the squared component perpendicular to the initial Bloch vector.
A small $|\mathbf q_j|$ means little branch-dependent motion; a
$\mathbf q_j$ parallel to $\mathbf r_j$ changes only the phase of the
conditional state without separating the two branches; a component of $\mathbf q_j$ perpendicular to $\mathbf r_j$ produces local branch separation.

For a fragment $\Frag$ and its complement $\Rem=\Env\setminus\Frag$, the product structure gives
\begin{equation}
\begin{aligned}
    B_{\Frag}(t)&=\prod_{j\in\Frag}B_j(t),
    &
    B_{\Rem}(t)&=\prod_{j\notin\Frag}B_j(t),
    \\
    B_{\Env}(t)&=B_{\Frag}(t)B_{\Rem}(t).
\end{aligned}
    \label{eq:methods_fragment_overlaps}
\end{equation}
Here $B_{\Frag}$ compares the two conditional states of the fragment,
whereas $B_{\Env}$ compares those of the full environment. Smaller overlaps
mean greater branch separation in the corresponding subsystem.
These overlaps play the same roles as the decoherence and macrofraction
distinguishability factors of spectrum broadcast structures, introduced
in
Refs.~\cite{KorbiczHorodeckiHorodecki2014Broadcasting,HorodeckiKorbiczHorodecki2015QuantumOrigins}
and used in dynamical studies of SBS formation
\cite{TuziemskiKorbicz2016BrownianSBS,MironowiczKorbiczHorodecki2017Monitoring}.
Because a fragment overlap is a product of local overlaps, we introduce two
single-witness exponents,
\begin{equation}
\begin{aligned}
    \kappa_{\rm typ}(t)&=-\mathbb E[\ln B_j(t)],
    \\
    \kappa_{\rm ann}(t)&=-\ln\mathbb E[B_j(t)].
\end{aligned}
    \label{eq:methods_overlap_exponents}
\end{equation}
The first describes a typical disorder realization, whereas the second
governs the disorder-averaged overlap.  For independent, identically
distributed (iid) witnesses,
$B_{\Frag}^{\rm typ}(t)\simeq e^{-|\Frag|\kappa_{\rm typ}(t)}$ and
$\mathbb E[B_{\Frag}(t)]=e^{-|\Frag|\kappa_{\rm ann}(t)}$, respectively
\cite{ZwolakRiedelZurek2014Chernoff}.
For the equal branch weights and pure conditional states considered here, these
overlaps determine the information diagnostics exactly.  Define
\begin{equation}
\begin{aligned}
    h(b)&=H_2\!\left(\frac{1+b}{2}\right),
    \\
    H_2(x)&=-x\log_2x-(1-x)\log_2(1-x).
\end{aligned}
    \label{eq:methods_binary_entropy}
\end{equation}
Then
\begin{equation}
\begin{aligned}
    \chi_z(\Ssys:\Frag)&=h(B_{\Frag}),
    \\
    I(\Ssys:\Frag)&=h(B_{\Env})+h(B_{\Frag})-h(B_{\Rem}),
    \\
    D_z(\Ssys:\Frag)&=h(B_{\Env})-h(B_{\Rem}).
\end{aligned}
    \label{eq:methods_overlap_information}
\end{equation}
Here $B_{\Frag}$ controls the pointer information in the observed fragment,
while $B_{\Rem}$ controls the coherence that remains after the unobserved
environment is traced out.  Without interactions within the environment, the
single-qubit branch dynamics therefore determine every fragment information
quantity exactly.
When both the fragment and its complement have small branch
overlaps, $\chi_z$ approaches $H_Z(\Ssys)$ and $D_z$ becomes small.
A broad range of such fragment sizes produces the classical
mutual-information plateau.

\subsection{Branch-conditioned dynamics}
\label{subsec:methods_branch_dynamics}

Equation~\eqref{eq:methods_branch_generators} gives the two evolutions of each
environment qubit conditioned on $s\in\{0,1\}$.  In the $s=0$ branch,
$P_1^{\Ssys}$ annihilates the interaction term and qubit $j$ evolves only under
its local field.  In the $s=1$ branch, it evolves under both the local field and
the C-INOT generator.  The scalar contribution
$(\pi g_j/2)I$ in $H_j^{(1)}$ produces only an overall phase of the
conditional branch state.  It does not affect branch-overlap magnitudes or the
entropic diagnostics considered here, so we remove it and write the effective
conditional Hamiltonians as
\begin{equation}
\begin{aligned}
    H_{0j}&=H_j^{(0)}=h_j\sigma_z,
    \\
    H_{1j}&=H_j^{(1)}-\frac{\pi g_j}{2}I=a_{xj}\sigma_x+a_{zj}\sigma_z,
\end{aligned}
    \label{eq:methods_conditional_hamiltonians}
\end{equation}
so that the branch propagators of Eq.~\eqref{eq:methods_branch_unitaries}
become
\begin{equation}
    U_j^{(0)}(t)=e^{-iH_{0j}t},
    \qquad
    U_j^{(1)}(t)=e^{-i\pi g_jt/2}\,e^{-iH_{1j}t}.
    \label{eq:methods_branch_unitaries_effective}
\end{equation}

The traceless branch-$1$ Hamiltonian defines a rotation vector in the $xz$
plane, with components
\begin{equation}
    a_{xj}=-\frac{\pi g_j}{2}\cos\theta,
    \qquad
    a_{zj}=h_j-\frac{\pi g_j}{2}\sin\theta.
    \label{eq:methods_branch_coefficients}
\end{equation}
Defining $\Omega_j=\sqrt{a_{xj}^2+a_{zj}^2}$ as the branch-$1$ rotation
frequency, the corresponding SU(2) propagator contains the phase $\Omega_jt$
and rotates about the time-independent axis
\begin{equation}
    \hat{\mathbf m}_j=(m_{xj},0,m_{zj})
    =\frac{(a_{xj},0,a_{zj})}{\Omega_j}.
    \label{eq:methods_branch_axis}
\end{equation}
Equivalently, the environment Bloch vector rotates through angle $2\Omega_jt$.
The branch-$0$ axis is $z$, with the analogous propagator phase $h_jt$.

To obtain the relative evolution explicitly, we introduce
\begin{align}
    c_h&=\cos(h_jt), & s_h&=\sin(h_jt),
    \\
    c_\Omega&=\cos(\Omega_jt), & s_\Omega&=\sin(\Omega_jt).
    \label{eq:methods_branch_trigonometric_factors}
\end{align}
Exponentiating the effective Hamiltonians in
Eq.~\eqref{eq:methods_conditional_hamiltonians} gives the SU(2) parts of the
branch propagators, which we continue to denote $U_j^{(s)}$ because the
omitted phase cancels from every overlap magnitude:
\begin{align}
    U_j^{(0)}(t)
    &=c_h I-i s_h\sigma_z,
    \\
    U_j^{(1)}(t)
    &=c_\Omega I-i s_\Omega
    \left(m_{xj}\sigma_x+m_{zj}\sigma_z\right).
    \label{eq:methods_conditional_propagators}
\end{align}
Although the axis $\hat{\mathbf m}_j$ is undefined when $\Omega_j=0$, the
propagator itself remains regular.  In this case we evaluate its axis-dependent
terms using the continuous limit
\begin{equation}
    s_\Omega m_{\mu j}
    =\frac{\sin(\Omega_jt)}{\Omega_j}a_{\mu j}
    \longrightarrow t a_{\mu j},
    \qquad \mu\in\{x,z\}.
    \label{eq:methods_zero_frequency_limit}
\end{equation}

Multiplying the two propagators and using
$\sigma_z\sigma_x=i\sigma_y$, we match
$U_j^{(0)\dagger}U_j^{(1)}$ to the SU(2) form in
Eq.~\eqref{eq:methods_relative_unitary}.  Its scalar and vector components are
\begin{align}
    q_{0j}&=c_hc_\Omega+s_hs_\Omega m_{zj},\\
    \mathbf q_j&=
    \begin{pmatrix}
        c_hs_\Omega m_{xj}\\
        -s_hs_\Omega m_{xj}\\
        c_hs_\Omega m_{zj}-s_hc_\Omega
    \end{pmatrix}.
    \label{eq:methods_relative_components}
\end{align}
The $y$ component is generated by the noncommutativity of the branch-$0$
$z$ rotation and the transverse part of the branch-$1$ rotation.  It vanishes
when either the local-field rotation or the transverse conditional rotation is
absent.

For the initial state $|\phi_j\rangle$ with Bloch vector $\mathbf r_j$, the
expectation value of the relative unitary is
\begin{equation}
    \langle\phi_j|U_j^{(0)\dagger}U_j^{(1)}|\phi_j\rangle
    =q_{0j}-i\mathbf q_j\cdot\mathbf r_j.
    \label{eq:methods_relative_expectation}
\end{equation}
Unitarity requires $q_{0j}^2+|\mathbf q_j|^2=1$, and the squared
branch-overlap magnitude becomes
\begin{equation}
    B_j^2=q_{0j}^2+(\mathbf q_j\cdot\mathbf r_j)^2
    =1-|\mathbf q_j\times\mathbf r_j|^2.
    \label{eq:methods_branch_overlap_reduction}
\end{equation}
This explicit branch evolution verifies the general identity in
Eq.~\eqref{eq:methods_record_separation}: only the component of the relative rotation perpendicular
to $\mathbf r_j$ separates the two conditional states.  The scalar branch phase
omitted from $H_{1j}$ changes the phase of their overlap but not its magnitude
$B_j$, so it leaves all fragment information diagnostics unchanged.

The following limits expose the roles of the self-field and the initial-state
alignment.  In the absence of a self-field, $h_j=0$,
Eq.~\eqref{eq:methods_branch_overlap_reduction}
reduces to
\begin{align}
    A_j(t)
    &=\Lambda(p,\theta)
    \sin^2\!\left(\frac{\pi g_jt}{2}\right),
    \\
    \Lambda(p,\theta)
    &=1-\left[\hat{\mathbf n}(\theta)\cdot\mathbf r(p)\right]^2,
    \qquad
    \hat{\mathbf n}=(\cos\theta,0,\sin\theta).
    \label{eq:methods_no_field_separation}
\end{align}
All dependence on $p$ and $\theta$ then enters through the geometric invariant
$\Lambda$ [Fig.~\ref{fig:methods_model}(b)].  At $p=1/2$ and $\theta=0$, one
has
$\mathbf r=\hat{\mathbf n}=\hat{\mathbf x}$, so $\Lambda=A_j=0$ and the
environment qubit acquires no local record of the system branch; we
call this preparation, with $\mathbf r\parallel\hat{\mathbf n}$, the aligned
point.

At the commuting-axis endpoint $\theta=\pi/2$, both conditional Hamiltonians
are diagonal in the $z$ basis.  Up to the omitted scalar phase, their relative
unitary is
\begin{equation}
    U_j^{(0)\dagger}U_j^{(1)}
    =\exp\!\left(+i\frac{\pi g_jt}{2}\sigma_z\right),
    \label{eq:methods_commuting_axis_control}
\end{equation}
which is independent of $h_j$.  A local $Z$ self-field therefore cannot alter
the record geometry at this endpoint.  Conversely, when $|h_j|\gg|g_j|$, both
conditional rotations are individually predominantly $z$-like: the branch-$0$
axis is exactly $\hat{\mathbf z}$, and the branch-$1$ axis $\hat{\mathbf m}_j$
is tilted away from $\hat{\mathbf z}$ only by
$|m_{xj}|=|a_{xj}|/\Omega_j=O(|g_j/h_j|)$.

We call the components of
$\mathbf q_j$ perpendicular to the field axis, $q_{xj}$ and $q_{yj}$,
transverse, and the component along the field axis, $q_{zj}$, longitudinal.
By Eq.~\eqref{eq:methods_relative_components}, both transverse components are
proportional to $m_{xj}$ and are suppressed by the same small ratio.  They
originate from the $\sigma_x$ part of the C-INOT generator, which does not
commute with the $\sigma_z$ self-field; for a strong field this noncommuting
part acts as a far-detuned drive and produces only a small, rapidly
oscillating transverse motion.  Strong fields consequently suppress
transverse, noncommuting record writing, but they need not suppress a record
written by the remaining longitudinal relative rotation.

The longitudinal component,
$q_{zj}=c_hs_\Omega m_{zj}-s_hc_\Omega$, is the relative phase accumulated
between the two nearly parallel $z$ rotations and reduces in this limit to
$\pm\sin[(\Omega_j-|h_j|)t]$.  The commuting $\sigma_z$ part of the C-INOT
generator shifts $\Omega_j-|h_j|$ at first order in $g_j$; at the aligned
point $\theta=0$ that part vanishes and only the second-order dispersive
detuning derived next remains.

In the strong-field limit at $p=1/2$, $\theta=0$, that remaining
rotation is slow but not absent.  The branch frequencies differ by the dispersive detuning
\begin{equation}
    \Delta\Omega_j
    =\sqrt{h_j^2+(\pi g_j/2)^2}-|h_j|
    \simeq \frac{\pi^2g_j^2}{8|h_j|},
    \qquad |h_j|\gg|g_j|.
    \label{eq:methods_dispersive_rate}
\end{equation}
The associated record-writing time is therefore
$t_{\rm disp}=O(|h_j|/g_j^2)$.  Strong-field suppression in the numerical
results is a finite-window statement: the transverse response is
off-resonant, while the residual longitudinal relative phase can write a
record at later times.

Finally, Eqs.~\eqref{eq:methods_relative_components} and
\eqref{eq:methods_branch_overlap_reduction} obey
$q_{0j}^2+|\mathbf q_j|^2=1$ and $0\le A_j\le1$.  We use these identities as
internal checks and compare the analytic relative unitary with direct
$2\times2$ matrix exponentiation at representative parameter points, modulo the
omitted scalar phase.

\subsection{Numerical protocol}
\label{subsec:methods_numerics}

Most reported information curves and maps are calculated from the exact
single-qubit propagators and fragment-overlap formulas.  These give
$I$, $\chi_z$, and $D_z$ directly for the pure product preparations and
noninteracting environment considered here.  We use them in
Figs.~\ref{fig:results_theta_dependence},
\ref{fig:results_holevo_plateau_overview}--\ref{fig:results_info_decomposition}
and \ref{fig:results_genericity}, the exact basis scan in
Fig.~\ref{fig:pointer_basis_test}, and Supplemental
Figs.~\ref{fig:supp_stability_endpoints}  and
\ref{fig:supp_theta_fields}.

We independently check these formulas by evolving the full state vector
with a second-order leapfrog integrator, using single-precision complex
amplitudes and $\Delta t=0.002$.  State-vector results in
Fig.~\ref{fig:results_scaling} and Supplemental
Figs.~\ref{fig:supp_matched_lambda_collapse} and
\ref{fig:supp_higher_n_validation} provide these checks.
The Supplemental Material also reports a numerical check with commuting
field and interaction axes.
Exact and state-vector comparisons use the same couplings, fragments,
and observation times.  Norm checks, time-step tests, and the treatment
of small information values are described in the Supplemental Material.

At each parameter point, we sample independent coupling and field
realizations and uniformly sample fragments of each size.  Comparisons
across preparations, angles, and field strengths reuse the same random
draws.  We calculate fragment means, quantiles, and record statistics
within each realization before averaging over realizations.
The 95\% confidence intervals resample whole realizations with replacement,
keeping their fragments fixed.  A separate check of uncertainty from
fragment sampling is given in the Supplemental Material.

An individual fragment meets our information criterion when
$\chi_z(\Ssys{:}\Frag)\geq(1-\delta)H_Z(\Ssys)$.
We define Holevo redundancy from the mean fragment information by
\begin{equation}
\begin{aligned}
    m_\delta(t)&=\min\left\{m:\overline{\chi_z(m,t)}
    \ge(1-\delta)H_Z(\Ssys)\right\},
    \\
    R_\delta(t)&=\frac{N_{\Env}}{m_\delta(t)},
\end{aligned}
    \label{eq:methods_redundancy}
\end{equation}
where $\overline{\chi_z(m,t)}$ is the mean over sampled fragments of size $m$
within one realization, so that $m_\delta$ is the smallest fragment size for
which an average fragment carries the fraction $1-\delta$ of the pointer
entropy in Holevo information. Substantial redundancy requires
$m_\delta\ll N_{\Env}$; a threshold crossing only at large fragment sizes
does not establish a broad plateau. As an additional check on broad fragment distributions, the
quantile-based statistic applies the same threshold to the $\alpha$ quantile
$q_\alpha[\chi_z(m,t)]$ over sampled fragments within one realization,
\begin{equation}
\begin{aligned}
    m_\delta^{(q_\alpha)}(t)
    &=\min\left\{m:q_\alpha[\chi_z(m,t)]
    \ge(1-\delta)H_Z(\Ssys)\right\},
    \\
    R_\delta^{(q_\alpha)}(t)
    &=\frac{N_{\Env}}{m_\delta^{(q_\alpha)}(t)}.
\end{aligned}
    \label{eq:methods_quantile_redundancy}
\end{equation}
Throughout, $\delta=0.1$.  We use $\alpha=0.1$ for the main quantile
criterion and also compare $\alpha=0.25$ in Supplemental
Fig.~\ref{fig:supp_percentile_comparison}.  The 10th-percentile criterion asks
that roughly $90\%$ of size-$m$ fragments carry at least $90\%$ of the
pointer-ensemble Holevo information.  When no evaluated size $m\leq N_{\Env}/2$ meets the
criterion, $R_\delta^{(q_\alpha)}$ is undefined rather than zero.

For time-dependent results, we take the median within each realization
over $t\in[20,40]$ for the no-field and parameter-sweep calculations,
and $t\in[30,60]$ for the field-profile comparison.  Redundancy medians
include only times with a threshold crossing.  These windows begin after
the no-field dephasing time, approximately $4.5$ time units; strong fields
can delay record formation beyond them.  The threshold-size comparisons use $t=36$; the system-basis scan uses
$t=60$.  All times use the unit $\hbar/E_0$ defined above.

To test whether records remain strong throughout an observation window,
we also calculate the minimum half-environment information in each realization:
\begin{equation}
    \chi_{\min}^{(k)}
    =\min_{t_i}
    \frac{q_{0.1}[\chi_z^{(k)}(m_h,t_i)]}{H_Z(\Ssys)},
    \qquad m_h=\frac{N_{\Env}}{2},
    \label{eq:methods_holevo_min}
\end{equation}
where $t_i$ runs over the sampled times in the chosen window and the
quantile is taken over sampled fragments in realization $k$.
We call that realization persistent when $\chi_{\min}^{(k)}\geq0.9$:
roughly $90\%$ of its sampled half-environment fragments then meet the
record threshold at every sampled time.  The passing fragments can change
with time, and this test makes no claim about unsampled fragments or times.
We report the mean of $\chi_{\min}^{(k)}$ over realizations; a mean above
$0.9$ does not imply that every realization passes.
This tests persistence of pointer information at half-environment
size. Redundancy and the classical plateau are assessed from the
fragment-size dependence of $\chi_z$ and $D_z$.

%% file: sections/figure_model_schematic.tex
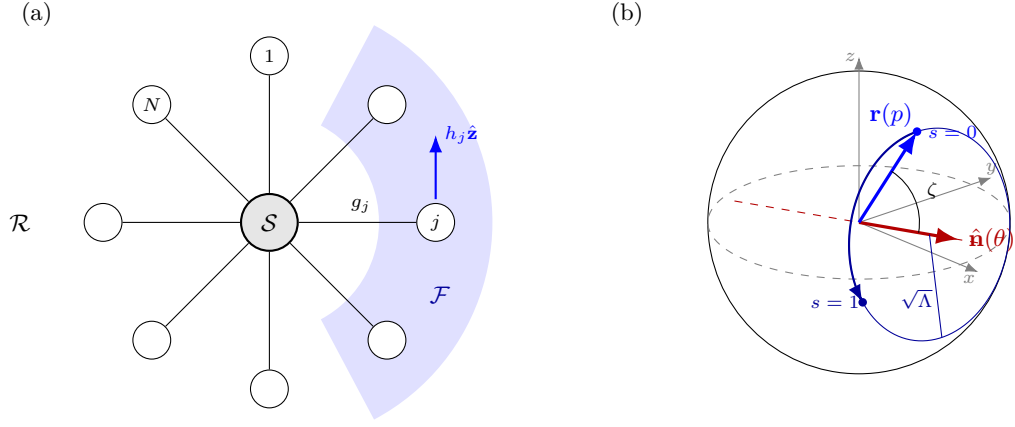
\begin{figure*}[t]
    \centering
    \begin{tikzpicture}[
        >=Latex,
        witness/.style={circle,draw,fill=white,minimum size=5mm,inner sep=0pt,font=\scriptsize},
        system/.style={circle,draw,thick,fill=gray!20,minimum size=7.5mm,inner sep=0pt,font=\small},
        every node/.style={font=\small}
    ]
    \begin{scope}
        \node[anchor=west] at (-3.4,2.75) {(a)};
        \fill[blue!12] (62:1.45) arc (62:-62:1.45) -- (-62:2.95) arc (-62:62:2.95) -- cycle;
        \node[blue!60!black] at (-23:2.45) {$\Frag$};
        \node at (180:3.3) {$\Rem$};
        \node[system] (S) at (0,0) {$\Ssys$};
        \foreach \ang/\name/\lab in {90/W1/{1},45/W2/{},0/W3/{$j$},-45/W4/{},-90/W5/{},-135/W6/{},180/W7/{},135/W8/{$N$}}{
            \node[witness] (\name) at (\ang:2.2) {\lab};
            \draw (S) -- (\name);
        }
        \node[font=\scriptsize] at ($(S)!0.55!(W3)+(0,0.22)$) {$g_j$};
        \draw[->,thick,blue] ($(W3.north)+(0,0.05)$) -- ++(0,0.85)
            node[right,font=\scriptsize] {$h_j\hat{\mathbf z}$};
    \end{scope}
    \begin{scope}[shift={(7.8,0)}]
        \node[anchor=west] at (-3.4,2.75) {(b)};
        \def\R{2.0}
        \pgfmathsetmacro{\pbeta}{35}    
        \pgfmathsetmacro{\ptheta}{10}   
        \pgfmathsetmacro{\Ad}{48}
        \pgfmathsetmacro{\Ae}{22}
        \pgfmathsetmacro{\uxx}{cos(\Ad)}
        \pgfmathsetmacro{\uxy}{sin(\Ad)}
        \pgfmathsetmacro{\wxx}{-sin(\Ae)*sin(\Ad)}
        \pgfmathsetmacro{\wxy}{sin(\Ae)*cos(\Ad)}
        \pgfmathsetmacro{\wxz}{cos(\Ae)}
        \draw[thin] (0,0) circle (\R);
        \draw[thin,dashed,gray] plot[domain=0:360,samples=120,variable=\t]
            ({\R*(\uxx*cos(\t)+\uxy*sin(\t))},{\R*(\wxx*cos(\t)+\wxy*sin(\t))});
        \draw[->,gray] (0,0) -- ({1.18*\R*\uxx},{1.18*\R*\wxx}) node[below left,font=\scriptsize,inner sep=1pt] {$x$};
        \draw[->,gray] (0,0) -- ({1.18*\R*\uxy},{1.18*\R*\wxy}) node[above,font=\scriptsize,inner sep=1pt] {$y$};
        \draw[->,gray] (0,0) -- (0,{1.18*\R*\wxz}) node[left,font=\scriptsize,inner sep=1pt] {$z$};
        \pgfmathsetmacro{\nX}{\R*\uxx*cos(\ptheta)}
        \pgfmathsetmacro{\nY}{\R*(\wxx*cos(\ptheta)+\wxz*sin(\ptheta))}
        \draw[dashed,red!70!black] ({-1.25*\nX},{-1.25*\nY}) -- ({1.25*\nX},{1.25*\nY});
        \draw[->,very thick,red!70!black] (0,0) -- (\nX,\nY) node[right] {$\hat{\mathbf n}(\theta)$};
        \pgfmathsetmacro{\rX}{\R*\uxx*sin(\pbeta)}
        \pgfmathsetmacro{\rY}{\R*(\wxx*sin(\pbeta)+\wxz*cos(\pbeta))}
        \coordinate (r) at (\rX,\rY);
        \draw[->,very thick,blue] (0,0) -- (r) node[above left,inner sep=1pt] {$\mathbf r(p)$};
        \pgfmathsetmacro{\angn}{atan2(\nY,\nX)}
        \pgfmathsetmacro{\angr}{atan2(\rY,\rX)}
        \draw[thin] (\angn:{0.4*\R}) arc (\angn:\angr:{0.4*\R});
        \node[font=\scriptsize,inner sep=1pt] at ({(\angn+\angr)/2}:{0.53*\R}) {$\zeta$};
        \pgfmathsetmacro{\sa}{sin(\pbeta+\ptheta)}
        \pgfmathsetmacro{\ca}{cos(\pbeta+\ptheta)}
        \pgfmathsetmacro{\px}{\sa*cos(\ptheta)}
        \pgfmathsetmacro{\pz}{\sa*sin(\ptheta)}
        \pgfmathsetmacro{\ux}{sin(\pbeta)-\px}
        \pgfmathsetmacro{\uz}{cos(\pbeta)-\pz}
        \draw[thin,blue!60!black] plot[domain=0:360,samples=120,variable=\t]
            ({\R*(\uxx*(\px+\ux*cos(\t)) - \uxy*\ca*sin(\t))},
             {\R*(\wxx*(\px+\ux*cos(\t)) - \wxy*\ca*sin(\t) + \wxz*(\pz+\uz*cos(\t)))});
        \draw[->,thick,blue!60!black] plot[domain=0:115,samples=60,variable=\t]
            ({\R*(\uxx*(\px+\ux*cos(\t)) - \uxy*\ca*sin(\t))},
             {\R*(\wxx*(\px+\ux*cos(\t)) - \wxy*\ca*sin(\t) + \wxz*(\pz+\uz*cos(\t)))});
        \coordinate (c) at ({\R*(\uxx*\px)},{\R*(\wxx*\px+\wxz*\pz)});
        \coordinate (rone) at
            ({\R*(\uxx*(\px+\ux*cos(115)) - \uxy*\ca*sin(115))},
             {\R*(\wxx*(\px+\ux*cos(115)) - \wxy*\ca*sin(115) + \wxz*(\pz+\uz*cos(115)))});
        \coordinate (rtwo) at ({\R*(\uxx*(\px-\ux))},{\R*(\wxx*(\px-\ux) + \wxz*(\pz-\uz))});
        \draw[thin,blue!60!black] (c) -- (rtwo)
            node[midway,anchor=north east,font=\scriptsize,inner sep=1pt] {$\sqrt{\Lambda}$};
        \fill[blue] (r) circle (1.6pt);
        \node[blue,anchor=west,xshift=3pt,font=\scriptsize,inner sep=1pt] at (r) {$s=0$};
        \fill[blue!60!black] (rone) circle (1.6pt);
        \node[blue!60!black,left,font=\scriptsize,inner sep=1pt] at (rone) {$s=1$};
    \end{scope}
    \end{tikzpicture}
    \caption{Model geometry.  (a) One system qubit $\Ssys$ coupled in a star
    geometry to $N\equiv N_{\Env}$ witness qubits with independent couplings
    $g_j$; each witness also experiences a local field $h_j$ along $z$,
    illustrated by the arrow at qubit $j$.  A fragment $\Frag$ (shaded) and its complement
    $\Rem=\Env\setminus\Frag$ partition the environment.  (b)
    Bloch-sphere geometry of one witness without a field.  The initial Bloch
    vector $\mathbf r(p)$ and the C-INOT axis $\hat{\mathbf n}(\theta)$ both
    lie in the $xz$ plane.  In the $s=0$ branch the witness stays at
    $\mathbf r$; in the $s=1$ branch it rotates about $\hat{\mathbf n}$ on the
    circle shown, whose radius $\sqrt{\Lambda}=\sin\zeta$ is set by the angle
    $\zeta$ between $\mathbf r$ and $\hat{\mathbf n}$.  The two conditional
    states separate only through this transverse component, so
    $\Lambda=1-(\hat{\mathbf n}\cdot\mathbf r)^2$ is the record-writing
    capacity of the witness, and an aligned preparation with
    $\mathbf r\parallel\hat{\mathbf n}$ writes no record.
    The relative evolution $U_j^{(0)\dagger}U_j^{(1)}$ compares
    the witness evolutions conditioned on the two system pointer values.
    The direction of $\mathbf q_j(t)$ gives the relative rotation axis,
    while its magnitude depends on the rotation angle.
    Without fields, this axis lies along $\hat{\mathbf n}$; local fields
    can change it.}
    \label{fig:methods_model}
\end{figure*}

%% file: sections/results.tex
\section{Results}
\label{sec:results}

\subsection{Static alignment controls record-writing capacity}
\label{subsec:results_static_alignment}

\begin{figure*}[!tp]
    \centering
    \includegraphics[width=\textwidth]{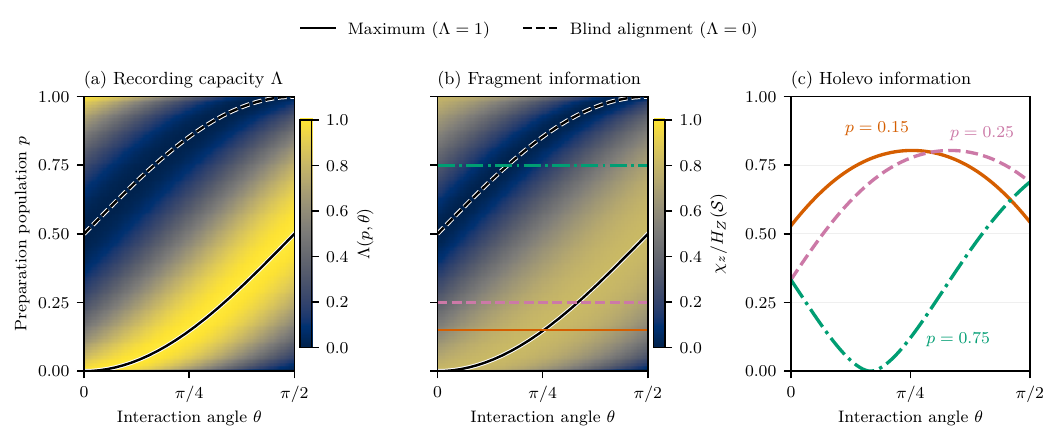}
    \caption{Preparation and interaction-angle dependence without local fields.
    (a) Alignment invariant $\Lambda$ over the full pure-preparation range $0\leq p\leq1$.
    (b) Corresponding Holevo information in fragments of size $m=2$, with
    $N_{\Env}=16$ and $g_j\sim\mathcal N(0,g^2)$, $g=0.1$.
    (c) Cuts through (b) at $p=0.15$, $0.25$, and $0.75$, marked by matching
    horizontal lines in (b); the vertical axis shows the same normalized
    Holevo information as the color scale in (b).
    Black solid curves in (a,b) mark maximal separation ($\Lambda=1$); black dashed curves mark
    blind alignment ($\Lambda=0$).  The lower-right corner,
    $p=0$, $\theta=\pi/2$, is also a blind preparation.
    We average over 64 sampled fragments, take the time median over
    $t=20,21,\ldots,40$ within each realization, and then average over
    24 realizations.  The same couplings and fragments are used across
    the grid.  The system
    starts in $\ket{+}$, so $H_Z(\Ssys)=1$ bit.}
    \label{fig:results_theta_dependence}
\end{figure*}

We first isolate static record writing by setting $W=h_0=0$.  Without a
field, the branch-$0$ propagator is the identity, and by
Eq.~\eqref{eq:methods_branch_coefficients} the relative branch unitary is a
rotation about the fixed C-INOT axis
$\hat{\mathbf n}(\theta)=(\cos\theta,0,\sin\theta)$ through the angle
$\pi g_jt$, so that $\mathbf q_j(t)=-\hat{\mathbf n}(\theta)\sin(\pi g_jt/2)$.
The preparation bias $p$ enters only through the initial Bloch vector
$\mathbf r(p)$, while the interaction angle $\theta$ enters only through the
axis $\hat{\mathbf n}(\theta)$.  Equation~\eqref{eq:methods_record_separation}
depends on them solely through $|\hat{\mathbf n}\times\mathbf r|^2$, the squared
sine of the angle between the conditional rotation axis and the witness's
initial state.  The full $(p,\theta)$ dependence therefore collapses onto a
single alignment invariant $\Lambda$: for every coupling realization,
Eqs.~\eqref{eq:methods_record_separation} and
\eqref{eq:methods_no_field_separation} give

    \begin{align}
        B_j^2(t)
        &=1-\Lambda(p,\theta)
        \sin^2\!\left(\frac{\pi g_jt}{2}\right),
        \\
        \Lambda(p,\theta)
        &=1-[\hat{\mathbf n}(\theta)\cdot\mathbf r(p)]^2,
        \qquad
        \hat{\mathbf n}=(\cos\theta,0,\sin\theta).
    \end{align}

Thus $\Lambda=0$ implies $B_j(t)=1$ at all times: the two conditional states
are locally indistinguishable and qubit $j$ acquires no record.  At
$\Lambda=1$ they can instead become orthogonal whenever the sine factor reaches
unity.  Intermediate values bound the attainable local branch separation.  The
invariant quantifies the record-writing capacity of one environment
qubit, but is not itself a redundancy or plateau measure.  This reduction is
the central simplification of the no-field problem: neither the gate angle
nor the preparation controls record writing on its own, and every diagnostic
built from branch overlaps must coincide for all $(p,\theta)$ pairs sharing
the same $\Lambda$.

Using the angular parameterization $p=\cos^2(\beta/2)$ and
$\mathbf r=(\sin\beta,0,\cos\beta)$, the invariant becomes
\begin{equation}
    \Lambda(\beta,\theta)=\cos^2(\beta+\theta).
\end{equation}
The conditional states are blind to the system branch when
\begin{equation}
    \beta+\theta=\frac{\pi}{2}\pmod{\pi}.
\end{equation}
On the principal blind-alignment curve this gives
\begin{equation}
    p_{\rm bad}(\theta)=\frac{1+\sin\theta}{2}.
\end{equation}
At the controlled-$Z$ endpoint, $\theta=\pi/2$, the south pole $p=0$ is
the second blind-alignment solution; both $z$-eigenstate preparations are then blind.  More
generally, no value of $\theta$ is intrinsically a good or bad witness.  As
the local equivalence in Eq.~\eqref{eq:methods_local_equivalence} makes clear,
the C-INOT ``imperfection'' rotates the environment-side conditional axis; the
record quality is fixed only after comparing that axis with the environment
preparation through $\Lambda(p,\theta)$.

Figure~\ref{fig:results_theta_dependence} shows two forms of nonmonotonic
dependence on $\theta$.  For $0<p<1/2$, fragment information first increases
as the interaction axis approaches orthogonality to the initial Bloch
vector, then decreases.  The cuts in panel (c) show that the optimal angle
depends on the initial environment preparation.  For $1/2<p<1$, it first falls to zero at blind
alignment and then recovers.  At fixed couplings and time,
fragment Holevo information increases with $\Lambda$, so this ordering also
survives fragment averaging, the time median within each realization,
and disorder averaging.  The monotonic
loss of information with gate angle for the $Z$-polarized environment of
Ref.~\cite{TouilYanGirolamiEtAl2022Eavesdropping} is therefore a
preparation-specific case.  The conventional CNOT imperfection parameter
alone does not determine how effectively an environment records the system.

We check that different preparations and interaction angles with the
same $\Lambda$ give the same information in environment fragments and the
same fragment sizes needed to reach the record threshold.  The numerical
results agree with this prediction from the exact overlap identity.
Supplemental Fig.~\ref{fig:supp_matched_lambda_collapse} gives the comparison
and sampling details.

The persistence measure of Eq.~\eqref{eq:methods_holevo_min} is
evaluated here for half-environment fragments ($m=8$) over $t\in[20,40]$.  The collapse
shows that the poor-record region at small $\Lambda$ reflects a failure of
conditional branch distinguishability, not weaker entangling dynamics or a
different system pointer observable.  This conclusion is restricted to pure product
preparations with $W=h_0=0$ and no environment--environment interaction.

We use the mean fragment Holevo information and its threshold
redundancy $R_{0.1}$, Eq.~\eqref{eq:methods_redundancy}, to characterize
record accumulation.  Lower fragment quantiles and late-window minima
provide additional robustness checks: they reveal fragments carrying
little information or temporary losses of recording that a mean or time median can conceal.  The
quantile-based $R_{0.1}^{(q_{0.1})}$ is such an additional diagnostic,
not a replacement for the mean-threshold definition.

\subsection{From local distinguishability to redundant macrofraction records}
\label{subsec:results_macrofraction_records}

\begin{figure}[!htbp]
    \centering
    \includegraphics[width=\columnwidth]{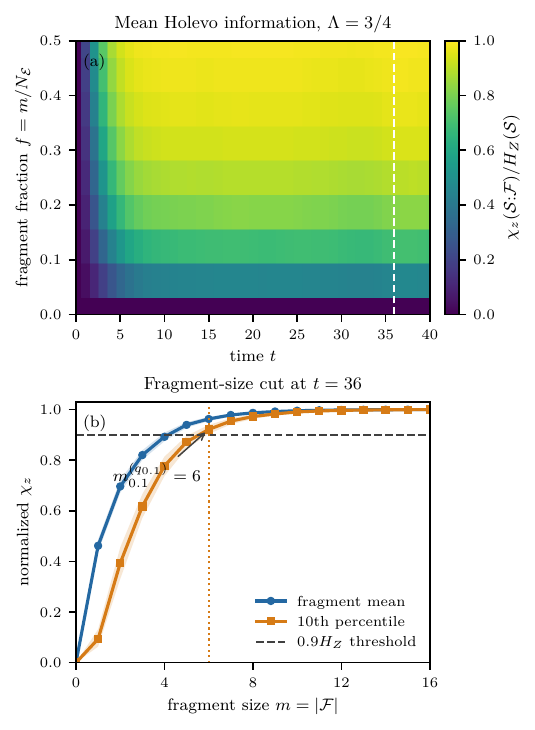}
    \caption{Accumulation of pointer-basis Holevo information for $N_{\Env}=16$,
$p=1/2$, $\theta=\pi/3$ ($\Lambda=3/4$), without self-fields or
environment--environment interactions.  (a) Fragment mean versus time and
$f=m/N_{\Env}$ for $m\leq8$, normalized by $H_Z(\Ssys)=1$ bit.
(b) Direct fragment-size cut at $t=36$ (dashed line in a), showing the
fragment mean and the mean within-realization 10th percentile.
Exact product overlaps are evaluated for up to 64 subsets per size in
each of 24 coupling realizations; all subsets are used when fewer exist.
Bands are 95\% confidence intervals obtained by resampling coupling realizations.  The mean curve first
crosses $0.9H_Z$ at $m=5$; the orange lower-quantile curve crosses at $m=6$.}
    \label{fig:results_holevo_plateau_overview}
\end{figure}

$\Lambda$ describes the record-writing capacity of one environment qubit.  A
nonzero value permits that qubit to distinguish the pointer branches, but QD
requires the same information to become encoded in many disjoint
fragments.  Because the environment qubits do not interact, the branch overlap
of a fixed fragment is an exact product.  Equivalently,
\begin{equation}
    K_{\Frag}(t)\equiv-\ln B_{\Frag}(t)
    =\sum_{j\in\Frag}\bigl[-\ln B_j(t)\bigr],
    \label{eq:results_fragment_log_overlap}
\end{equation}
so that $K_{\Frag}$ is the additive log-overlap accumulated from the witnesses in
$\Frag$.  Random couplings do not alter this identity, but they generate a
distribution of $K_{\Frag}$ across fragment choices and realizations, so the
typical fragment, the average fragment, and the lower tail of the fragment
distribution can behave differently.

The typical and disorder-averaged exponents of
Eq.~\eqref{eq:methods_overlap_exponents} summarize two corresponding averages,
\begin{align}
    B_{\Frag}^{\rm typ}(m,t)&\simeq e^{-m\kappa_{\rm typ}(t)},
    \\
    \mathbb E[B_{\Frag}(m,t)]&=e^{-m\kappa_{\rm ann}(t)},
    \qquad m=|\Frag|.
    \label{eq:results_fragment_exponents}
\end{align}
A separation between $\kappa_{\rm typ}$ and $\kappa_{\rm ann}$ therefore
signals a broad distribution of fragment quality.  For the equal-weight pure branches used
here, the same overlap fixes both the pointer-basis Holevo information and the
optimal binary discrimination error,
\begin{equation}
    \chi_z(\Ssys{:}\Frag)=h(B_{\Frag}),
    \qquad
    P_{\rm err}^{\rm opt}(\Frag)
    =\frac{1-\sqrt{1-B_{\Frag}^2}}{2}.
    \label{eq:results_fragment_discrimination}
\end{equation}
Thus reducing $B_{\Frag}$ multiplicatively with fragment size simultaneously
drives $\chi_z$ toward $H_Z(\Ssys)$ and the minimum
single-shot branch-discrimination error toward zero
~\cite{Helstrom1976QuantumDetection}.
For this equiprobable binary pure-state ensemble, the Helstrom
measurement also attains the accessible information~\cite{Levitin1995OptimalMeasurements},
giving
\begin{equation}
    I_{\rm acc}(\Ssys{:}\Frag)
    =1-H_2\!\left(P_{\rm err}^{\rm opt}(\Frag)\right)
    \leq \chi_z(\Ssys{:}\Frag).
    \label{eq:results_accessible_information}
\end{equation}
Our redundancy thresholds are defined using the Holevo quantity $\chi_z$;
$P_{\rm err}^{\rm opt}$ provides the corresponding operational
branch-discrimination diagnostic
\cite{TouilYanGirolamiEtAl2022Eavesdropping}.
For an individual fragment in this pure-state ensemble, the $90\%$ Holevo threshold implies an optimal
branch-discrimination error of at most about $3.5\%$.  The accessible
information at the threshold is about $0.78$ bits, rather than $0.9$ bits.

Figure~\ref{fig:results_holevo_plateau_overview} illustrates how imperfect
local records accumulate into a fragment record at $\Lambda=3/4$.
The mean Holevo information develops a broad plateau and exceeds $0.9H_Z$
at five qubits for $t=36$, well before the full environment is accessed.
The orange curve checks fragments carrying less information: we take the 10th percentile
in each realization and then average those values.  Its later crossing
shows the extra fragment size needed to make the record less sensitive to
fragment choice.  Thus a high-information plateau does not require every
individual environment qubit to be an ideal witness.

\begin{figure*}[tp]
    \centering
    \includegraphics[width=\textwidth]{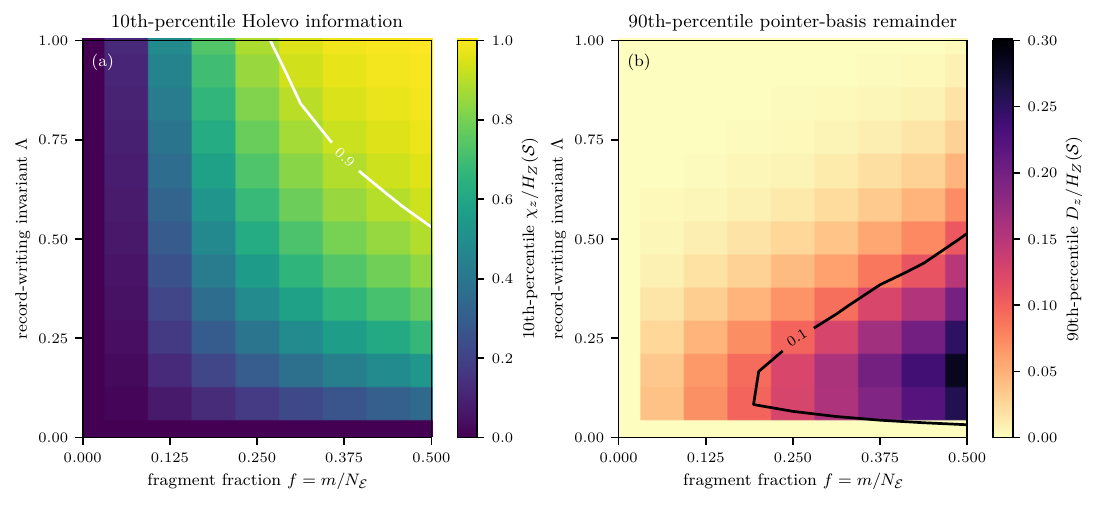}
    \caption{Fragment statistics versus $\Lambda$ and $f=m/N_{\Env}$ for $N_{\Env}=16$
without self-fields or environment--environment interactions.
(a) Within-realization 10th percentile of $\chi_z/H_Z$.
(b) Within-realization 90th percentile of the pointer-basis remainder
$D_z/H_Z$, where $D_z=I-\chi_z$.  Both panels use exact local propagators.  Up to 64 subsets are sampled per size,
with all 16 single-site subsets used at $m=1$.
Each statistic is summarized by its median over $t\in[20,40]$;
preparations with the same $\Lambda$ are averaged within each of 24 paired coupling
realizations, followed by the disorder mean.  White and black contours mark
the resulting $0.9$ and $0.1$ levels, respectively.
A value of one denotes full pointer-ensemble Holevo information only in
(a).}
    \label{fig:results_fragment_sweep}
\end{figure*}

Figure~\ref{fig:results_fragment_sweep} brings two aspects of
record quality together.  The Holevo information tests how much
of the pointer value a fragment can encode; the remainder $D_z$ tests how
much of its correlation with the system lies outside that record.
The region with high Holevo information and a small remainder supports
predominantly classical records.  As $\Lambda$ decreases, larger fragments
are needed to accumulate enough information.  Near the blind boundary,
both quantities become small: these fragments have little quantum
correlation because they have acquired little information at all.
Conversely, an appreciable remainder marks correlations that have not been
converted into an effectively classical pointer record.

The lower Holevo percentile and upper remainder percentile test
fragments carrying less pointer information or more correlation outside
the pointer record, respectively. Their contours locate the
regions satisfying each condition after taking time medians and disorder averages;
they do not count fragments passing both conditions simultaneously.
The maps thus connect the local alignment mechanism to the fragment sizes
and correlation structure of the resulting records.
Systematic dependence on $N_{\Env}$ is considered in
Sec.~\ref{subsec:results_macrofraction_scaling}.

\subsection{Self-evolution changes the relative branch rotation}
\label{subsec:results_self_evolution}

Local self-evolution changes record formation through the relative branch
rotation, rather than through the field strength alone.  The exact one-qubit
result remains
\begin{equation}
    A_j(t)=|\mathbf q_j(t)\times\mathbf r_j|^2,
\end{equation}
but the rotation vector $\mathbf q_j(t)$ now depends on the local
field $h_j$.  A field can therefore rescue a blind witness by moving
$\mathbf q_j$ away from the initial Bloch vector $\mathbf r_j$, or degrade an
effective witness by making the two more nearly parallel.  At large $|h_j|$,
it can also delay record writing beyond the observation window because the
branch dynamics becomes predominantly $z$-like and the transverse,
noncommuting motion is off-resonant.  The residual response at the aligned
point then occurs on the
dispersive timescale $t_{\rm disp}=O(|h_j|/g_j^2)$ from
Eq.~\eqref{eq:methods_dispersive_rate}.

\begin{figure*}[tp]
    \centering
    \includegraphics[width=\textwidth]{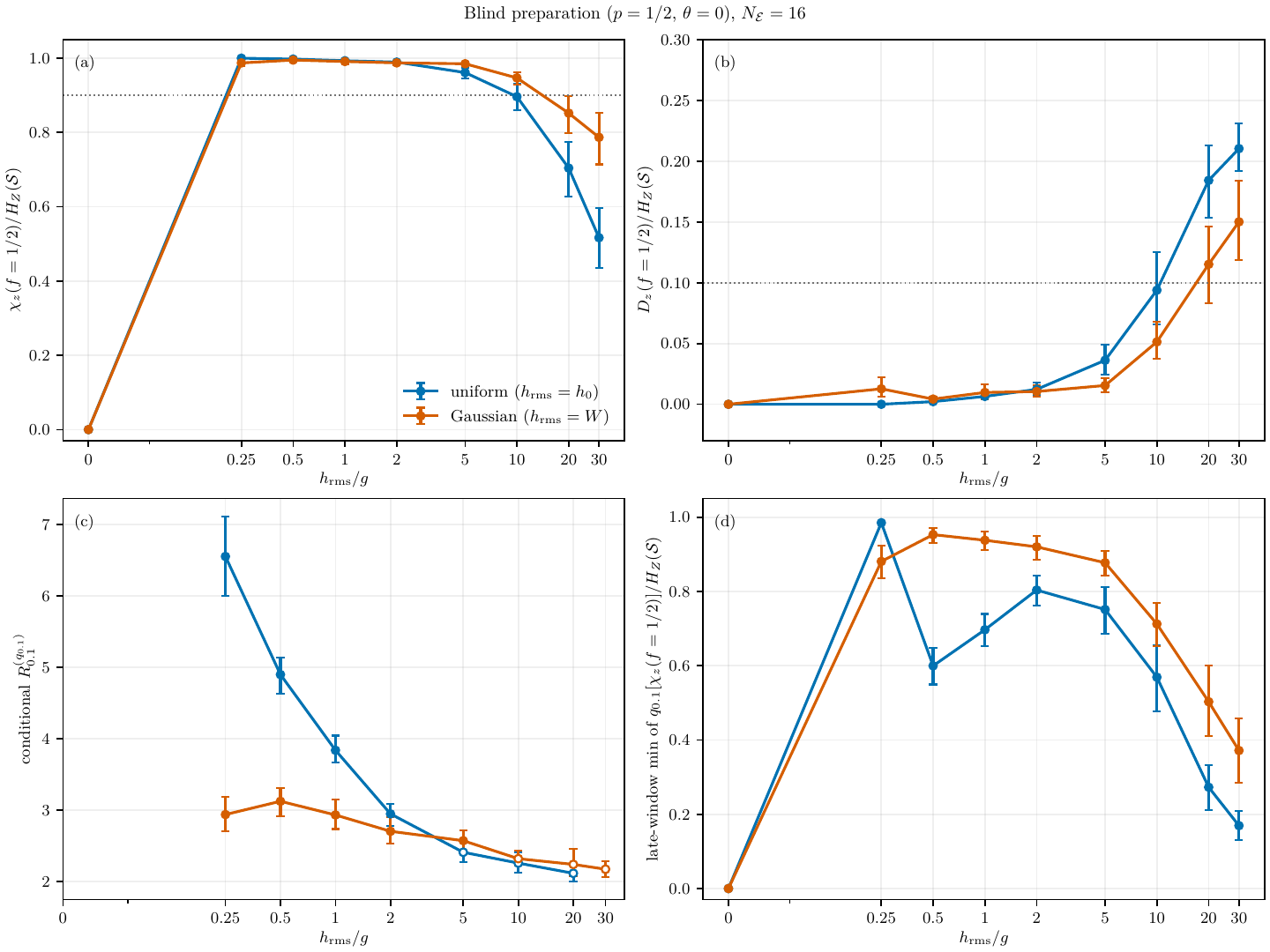}
    \caption{Uniform (blue) and Gaussian (orange) $Z$ fields at the blind preparation
$p=1/2$, $\theta=0$, with $N_{\Env}=16$ and 24 realizations per point.
The horizontal axis is $h_{\rm rms}/g$, where $h_{\rm rms}=h_0$ or $W$;
its symmetric-logarithmic scale resolves the small-field region.
(a,b) Fragment-mean half-environment Holevo information and pointer-basis
remainder, divided by $H_Z(\Ssys)$.  (c) Quantile-based redundancy
$R_{0.1}^{(q_{0.1})}$, conditional on a threshold crossing
at an evaluated time.  Open symbols indicate fewer than 24 defined
realizations: 21, 15, 5 for uniform fields at $5g$, $10g$, $20g$, and
8, 21, 22, 20, 11, 5 for Gaussian fields at $0.10g$, $0.15g$, $0.20g$,
$10g$, $20g$, $30g$, respectively.
No point is shown when no realization crosses.
(d) Mean of the realization-level time minima,
Eq.~\eqref{eq:methods_holevo_min}.  Panels (a--c) average time medians
across realizations.  All time summaries use $t\in[30,60]$;
bars are 95\% intervals over realizations.
Sensitivity to fragment sampling is reported in the
Supplemental Material; the bars shown here condition on the sampled subsets.}
    \label{fig:results_self_evolution_map}
\end{figure*}

We isolate this mechanism at the exactly blind point $p=1/2$, $\theta=0$ by
comparing uniform and zero-mean Gaussian $Z$-field profiles at the same root-mean-square
(rms) field strength $h_{\rm rms}$, equal to $h_0$ for the uniform
profile and to $W$ for the Gaussian one, in units of $g\equiv\sigma_g=0.1$.
Figure~\ref{fig:results_self_evolution_map} shows how weak self-fields
restore records at this initially blind preparation.  Near the onset,
records carry substantial information but weaken during the observation
window.  Persistent records appear at somewhat stronger fields, with
uniform fields establishing persistence more consistently in this range.
Additional cases at $h_{\rm rms}\leq0.2g$ and their numerical
checks are given in the Supplemental Material.

At intermediate strengths, uniform fields can produce high time-median
information with pronounced dips.  They give all witnesses the same
branch-$0$ rotation, although the branch-$1$ frequencies
$\Omega_j=\sqrt{h_0^2+(\pi g_j/2)^2}$ still differ through the couplings.
Random fields add phase dispersion and weaken coherent revivals.
They maintain high lower-percentile information over a broader range of
sampled strengths, without ensuring persistence in every realization.

The spread of field strengths also helps explain the crossover between
profiles.  Where a field helps recording, a uniform profile gives every
witness that strength; a Gaussian profile includes weaker or less favorable
fields.  Where strong fields delay recording, the Gaussian profile includes
witnesses with smaller fields that can still write records.  The spread
therefore broadens both enhancement and suppression.  Which profile performs
better depends on the information statistic and observation window, with
revival patterns also contributing.  At the strongest fields, both profiles
lose effective recording within the window, but the Gaussian decline is
more gradual.
Interpolating between the two profiles at fixed rms strength
(Supplemental Fig.~\ref{fig:supp_field_mean_width}) shows that a modest
spread of local fields already removes the intermediate-field dips of the
uniform profile, without requiring a zero mean.

Self-fields also change the dependence on interaction angle.
At $p=1/2$, where the no-field information rises monotonically with $\theta$
(Fig.~\ref{fig:results_theta_dependence}), a uniform field can make an
intermediate angle yield more time-median fragment information than either
endpoint (Supplemental Fig.~\ref{fig:supp_theta_fields}). This maximum is
less pronounced for Gaussian-distributed fields at the same rms strength.

The same field-strength sweep becomes a null result when the axes commute.  At
$p=1/2$ and $\theta=\pi/2$, neither uniform nor Gaussian $Z$ fields up to $30g$
measurably change the late-window Holevo information, pointer-basis remainder,
or persistence measure (see the numerical check in the Supplemental
Material).  Thus comparable field
strengths affect the record only when the field and conditional record-writing
axes do not commute.  The response in
Fig.~\ref{fig:results_self_evolution_map} therefore arises from the
noncommuting field and record-writing axes, not from field strength alone.

\subsection{Fragment information and the pointer-basis remainder}
\label{subsec:results_holevo_discord}

\begin{figure*}[tp]
    \centering
    \includegraphics[width=\textwidth]{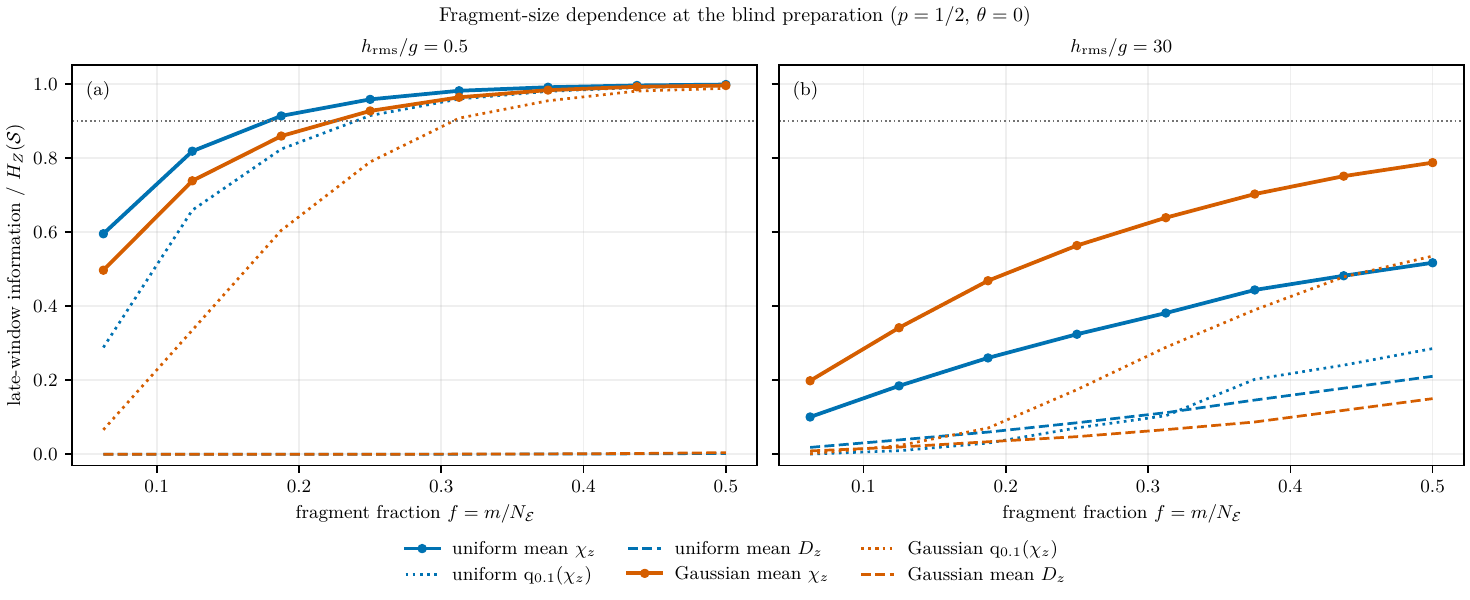}
    \caption{Fragment-size dependence at $p=1/2$, $\theta=0$, $N_{\Env}=16$,
for uniform (blue) and Gaussian (orange) $Z$ fields.
Solid, dotted, and dashed curves denote the fragment mean Holevo
information, within-realization 10th percentile, and mean pointer-basis
remainder, respectively, normalized by $H_Z(\Ssys)$.
Each curve is the disorder mean of realization-level medians over
$t\in[30,60]$, using 24 realizations and up to 64 subsets per size.
(a) $h_{\rm rms}=0.5g$.  (b) $h_{\rm rms}=30g$.
All curves use the exact product-overlap formulas.}
    \label{fig:results_info_decomposition}
\end{figure*}

Figure~\ref{fig:results_info_decomposition} compares the mean and
10th-percentile Holevo information with the pointer-basis remainder at
moderate and strong fields.  The separation between the mean and the
lower percentile shows how a high mean can conceal fragments carrying
little information.  As in Sec.~\ref{sec:introduction}, a small
remainder is informative only together with high Holevo information at the
same fragment size; this joint assessment excludes the blind zero-field
boundary, where both vanish.

\subsection{Exact fragment thresholds and finite-size validation}
\label{subsec:results_macrofraction_scaling}

The fragment product in Eq.~\eqref{eq:methods_fragment_overlaps} lets us
calculate thresholds for large environments without evolving a state vector.
At fixed $t$ and $\Lambda$, independent couplings give independent local
overlaps, so the negative log overlap of a typical fragment grows linearly
with its size,
\begin{equation}
    -\ln B_{\Frag}^{\rm typ}(m,t)\simeq m\kappa_{\rm typ}(t),
    \qquad m=|\Frag|.
    \label{eq:results_typical_overlap_scaling}
\end{equation}
Let $b_\delta$ solve $h(b_\delta)=1-\delta$.  Replacing the fragment overlap by
its typical value gives the continuous estimate
\begin{equation}
    m_\delta^{\rm typ}
    =\frac{\ln(1/b_\delta)}{\kappa_{\rm typ}}.
    \label{eq:results_threshold_estimate}
\end{equation}
For $\delta=0.1$, $b_{0.1}=0.36796$.  This estimate captures the inverse
$\kappa_{\rm typ}$ dependence.  However, evaluating $h$ at the typical overlap
is not equivalent to averaging $h(B_{\Frag})$, and the estimate does not
describe the lower tail of the fragment distribution.

\begin{figure*}[tp]
    \centering
    \includegraphics[width=\textwidth]{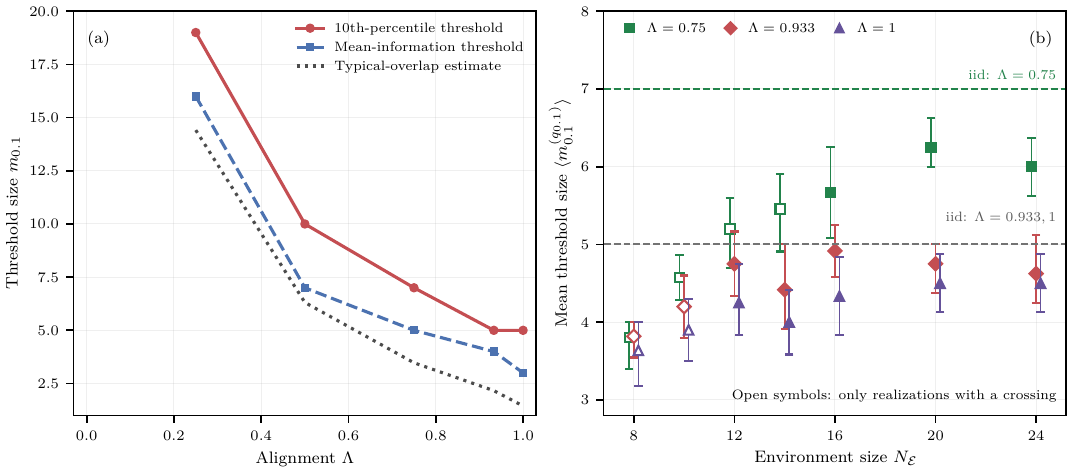}
    \caption{Fragment thresholds at $t=36$ without self-fields or
environment--environment interactions.  (a) Threshold size $m_{0.1}$ for
$90\%$ of the pointer-ensemble Holevo information, applied to the mean
(blue) or the 10th percentile (red), from $5\times10^5$ iid coupling
fragments per $\Lambda$.  Connecting lines are guides between sampled
alignments; integer thresholds need not vary smoothly.
The iid model fixes no total $N_{\Env}$.  At $\Lambda=0$ no finite
threshold exists.  The dotted curve is the continuous typical-overlap
estimate $\ln(1/b_{0.1})/\kappa_{\rm typ}$.
(b) Mean threshold size $\langle m_{0.1}^{(q_{0.1})}\rangle$,
found separately in each realization before averaging.
Symbols show state-vector results; exact calculations give the same values.
Filled symbols include all realizations; open symbols include only those
with a crossing among $m\leq N_{\Env}/2$.  Crossing counts and sampling
details are given in the Supplemental Material.  Bars are 95\% confidence
intervals obtained by resampling realizations of the exact calculation.
Dashed lines show the iid thresholds from (a), with the same value for
$\Lambda=0.933$ and $1$.  Small horizontal offsets separate coincident symbols.
All three alignments are sampled through $N_{\Env}=24$.}
    \label{fig:results_scaling}
\end{figure*}

We therefore calculate the threshold directly from the exact local formula
\begin{equation}
    B_j=\sqrt{1-\Lambda\sin^2(\pi g_jt/2)}
\end{equation}
by sampling $5\times10^5$ independent coupling fragments at $t=36$,
with $g_j\sim\mathcal N(0,0.1^2)$.  Figure~\ref{fig:results_scaling}(a)
shows that increasing $\Lambda$ reduces the fragment size needed to reach the
mean Holevo threshold.  The lower-quantile threshold is larger, exposing
the extra fragment size needed for reliable recording across
fragment choices.
The typical-overlap estimate captures the trend but underestimates these
thresholds because it omits the distribution of fragment overlaps.

Figure~\ref{fig:results_scaling}(b) shows the required fragment size
directly: we find the quantile threshold in each realization and then
average the sizes with a crossing.  At the largest sampled environments,
these sizes vary little, consistent with a record requiring a roughly fixed
number of qubits.  Open symbols at smaller sizes describe only realizations
that cross within the evaluated half environment and therefore favor smaller
thresholds.  Exact and state-vector calculations give identical thresholds
for the same couplings and sampled fragments.
The horizontal references use the iid fragment distribution from panel (a).
Averaging thresholds obtained from finite environments is a different
statistic, so these references need not coincide with the plotted means.
Using mean information or the 25th percentile instead of the 10th percentile
reduces the required size while preserving its weak dependence on environment
size at the larger sampled sizes (Supplemental
Fig.~\ref{fig:supp_percentile_comparison}).  We retain the 10th percentile
to test reliability across fragment choices.  The finite environments
therefore demonstrate reliable records with modest redundancy, while the
fixed-size prediction describes how their multiplicity grows.
In the iid model, adding qubits outside a fragment leaves its overlap
distribution unchanged.  The resulting fixed threshold gives linear growth
of redundancy with $N_{\Env}$.

As a check with larger environments, we repeated the no-field
$\Lambda=1/2$ case and four aligned-field cases at $N_{\Env}=20$, using the
same eight realization seeds as the corresponding $N_{\Env}=16$ runs.  The two
weak-field rescue cases retain near-unit pointer information and
a disorder-averaged late-window minimum above the record threshold,
whereas neither $30g$ field profile
produces a robust record.  The no-field curves also show the same
multiplicative growth of information with fragment size.  Supplemental
Fig.~\ref{fig:supp_higher_n_validation} documents these five parameter checks.

A further fixed-time check of the same no-field case at $N_{\Env}=24$
shows increasing half-environment Holevo information and decreasing
pointer-basis remainder with environment size.  This supports the
product-overlap explanation of stronger fragment records in larger
environments.  The matched-seed comparison and numerical values are given
in the Supplemental Material.

\subsection{High fragment information over a broad parameter region}
\label{subsec:results_genericity}

\begin{figure*}[tp]
    \centering
    \includegraphics[width=\textwidth]{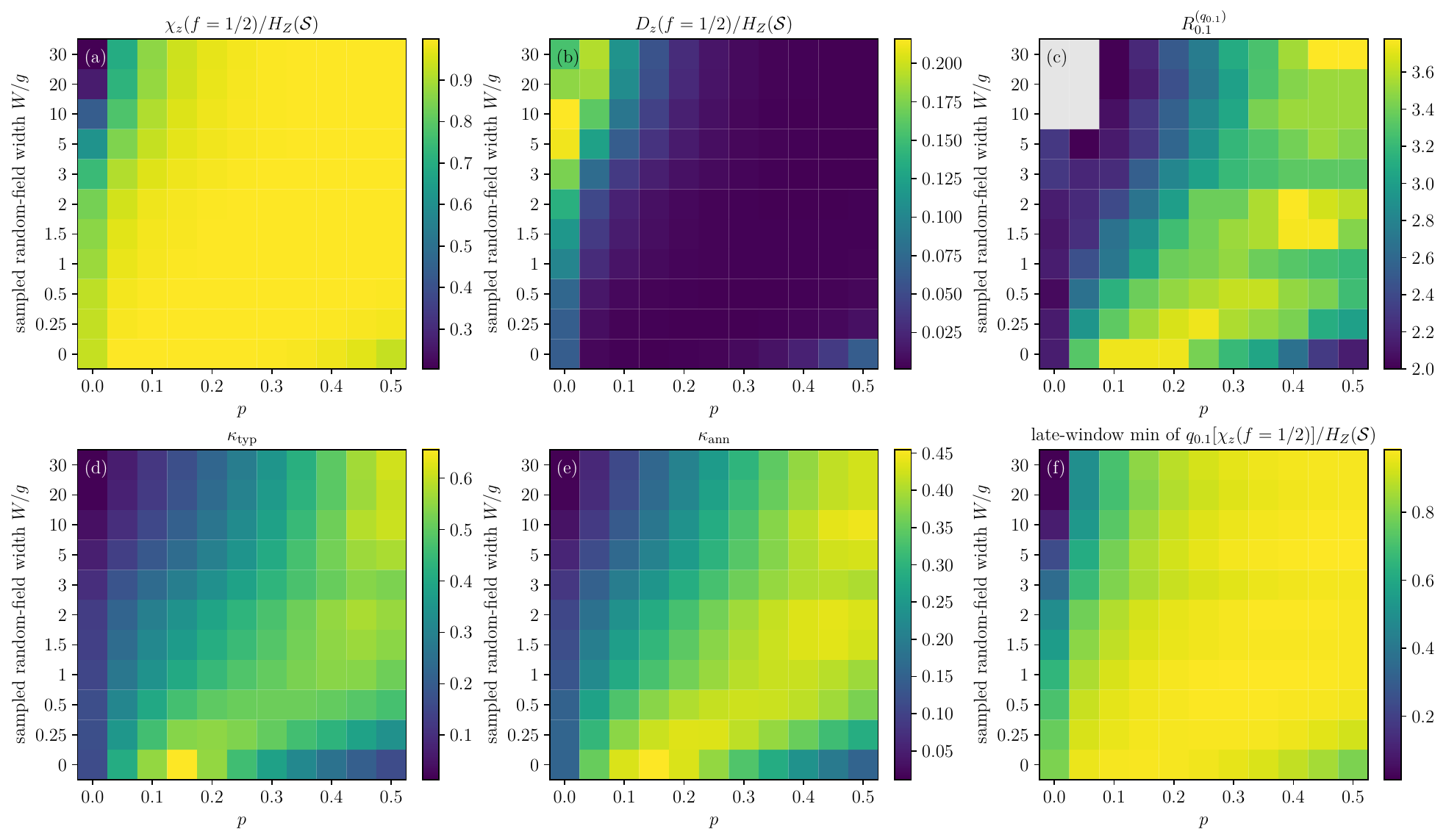}
    \caption{Parameter sweep using exact overlaps at $\theta=\pi/4$, $p\in[0,0.5]$, with
$N_{\Env}=16$, 12 realizations per point, and no environment--environment
interaction.  Rows are categorical, labeled by
$W/g\in\{0,0.25,0.5,1,1.5,2,3,5,10,20,30\}$; cell area is not a
parameter-space area measure.  The top row shows normalized
half-environment Holevo information, pointer-basis remainder, and
quantile-based redundancy $R_{0.1}^{(q_{0.1})}$.  The bottom row shows typical and
disorder-averaged overlap exponents, and the disorder mean of the
realization-level minimum in Eq.~\eqref{eq:methods_holevo_min},
for $m=8$ and sampled times in $[20,40]$.  The other panels average
realization-level time medians; redundancy medians include only times
with a threshold crossing.
The mean minimum is at least $0.9$ at 85 of 121 points; all 12
realizations individually pass at 22 points.  Gray redundancy cells
have no threshold crossing within $m\leq8$.  The no-field blind point
$p\simeq0.854$ lies outside the plotted range.}
    \label{fig:results_genericity}
\end{figure*}

The preparation and field sweeps show that effective record writing
occupies a broad regime rather than a finely tuned parameter line
(Fig.~\ref{fig:results_genericity} and Supplemental
Fig.~\ref{fig:supp_stability_endpoints}).  At $\theta=\pi/4$, much of the
sampled preparation range already separates the conditional branches
without a field; weak and moderate fields retain high fragment information
across an extended region.  At stronger fields, the record becomes more
preparation dependent, consistent with a predominantly $Z$-directed
conditional rotation: a $Z$ eigenstate is insensitive to that rotation,
whereas a state with transverse Bloch components can retain a record
through relative phases.  Record quality is therefore controlled by the
preparation and relative branch geometry, rather than field strength alone.
The displayed slice excludes the
no-field blind-alignment curve, so it does not map the full boundary of
this regime.

The no-field geometry also shows that blind alignments are atypical for
randomly oriented pure witnesses.  If each environment qubit is prepared
independently in a Haar-random pure state, then
isotropy of its Bloch vector gives
\begin{equation}
    \mathbb E_{\rm Haar}\!\left[(\hat{\mathbf n}\cdot\mathbf r)^2\right]
    =\frac{1}{3},
    \qquad
    \mathbb E_{\rm Haar}[\Lambda]=\frac{2}{3}.
\end{equation}
Independent random product witnesses are therefore not generically trapped on
the blind-alignment curve, consistent with the broader emergence of a common
classical observable for many environmental observers
\cite{BrandaoPianiHorodecki2015Generic}.  This statement concerns a product of
independently sampled single-qubit states; it does not extend to a globally
Haar-random, entangled environment state.

%% file: sections/discussion.tex
\section{Discussion}
\label{sec:discussion}

\subsection{Physical mechanism}
\label{subsec:discussion_mechanism}

Within the imperfect-CNOT model with local environment fields, the emergence of a
redundant pointer record is controlled by the distinguishability of the
conditional fragment states.  The instantaneous branch separation of witness $j$ is
fixed by its relative branch geometry,
\begin{equation}
    A_j(t)=|\mathbf q_j(t)\times\mathbf r_j|^2.
\end{equation}
This quantity separates the nonlocal content of the controlled gate from the
capacity of a particular environment preparation to register its action.  As
shown in Eq.~\eqref{eq:methods_local_equivalence}, every C-INOT gate is obtained
from CNOT by conjugating the environment qubit with local $y$ rotations.  It
therefore has the same entangling content.  Nevertheless, a witness
initially aligned with its conditional rotation has $A_j=0$ and acquires no
branch information.

This construction also applies to other models in which the system
controls a unitary on an environment qubit.  Removing its scalar phase
always gives a relative rotation $q_0I-i\mathbf q\cdot\boldsymbol\sigma$;
the component perpendicular to the initial Bloch vector determines branch
separation.  Our no-field $\Lambda$ is the corresponding squared perpendicular
component of the preparation relative to a fixed rotation axis.  With more
general dynamics, the relative axis itself can change in time.  The
geometric test still applies, although a single time-independent $\Lambda$
need not describe the whole evolution.

The local field changes the relative branch rotation $\mathbf q_j(t)$, not
the system observable selected by the interaction.  Self-evolution can
therefore rescue, degrade, or delay record writing according to whether it
creates or removes a component of $\mathbf q_j$ perpendicular to
$\mathbf r_j$.  This also explains why the effect is nonmonotonic in field
strength.  The oscillatory time dependence comes from the phases $h_jt$ and
$\Omega_jt$ in the conditional rotations.  Strong fields
make the transverse, noncommuting contribution off-resonant, rather than
eliminating every possible record.  For the aligned preparation, the residual
branch phase accumulates at
$\Delta\Omega_j\simeq\pi^2g_j^2/(8|h_j|)$, so the poor-record region at large
$|h_j|$ is explicitly a statement about a finite observation window.

\subsection{Relation to previous work}
\label{subsec:discussion_mironowicz}

Our analysis applies the distinguishability approach of earlier
spin-environment studies
\cite{ZwolakRiedelZurek2016SpinEnvironments} to the C-INOT gate
family \cite{TouilYanGirolamiEtAl2022Eavesdropping,MironowiczHorodeckiHorodecki2022SelfEvolution}.
We resolve its preparation and field dependence, compare field distributions
at equal root-mean-square strength, and quantify reliability across fragments.
Relative to the few-qubit analysis of Mironowicz \emph{et al.}
(Sec.~\ref{sec:introduction}), we isolate the pure-product preparation and
test the consequences of that geometry for larger fragments.

The same conditional-axis combination also appears in the analytic variables
of Ref.~\cite{MironowiczHorodeckiHorodecki2022SelfEvolution}.  In its
noninteracting, unit-coupling sector, the quantity denoted there by $y$ is
$y=-2a_z$, with $a_z=h_j-(\pi g_j/2)\sin\theta$ in our notation.  This
coefficient vanishes at
\begin{equation}
    h_j^{\rm crit}=\frac{\pi g_j}{2}\sin\theta,
    \label{eq:discussion_critical_field}
\end{equation}
where the branch-$1$ rotation axis becomes purely transverse.  Record
quality, however, is set by the full relative rotation
$U_j^{(0)\dagger}U_j^{(1)}$ and by its component perpendicular to
$\mathbf r_j$: canceling $a_{zj}$ leaves the branch-$0$ rotation in place, so
the local branch separation at $h_j^{\rm crit}$ still depends on $p$, $\theta$,
$g_j$, and time through Eq.~\eqref{eq:methods_relative_components}.  The relative-rotation variables thus identify a common dynamical ingredient.
This does not reproduce all of their few-qubit extrema: their SBS-distance
optimization and mixed initial witnesses address a broader problem than
the pure-state fragment diagnostics used in our large-environment calculations.

The two papers also use superficially similar preparation parameters for
different physical quantities.  Our parameter $p\in[0,1]$ fixes the coherent
pure state $\sqrt{p}\ket{0}+\sqrt{1-p}\ket{1}$ and rotates a unit Bloch vector
through the $xz$ plane.  By contrast, Ref.~\cite{MironowiczHorodeckiHorodecki2022SelfEvolution}
uses
\begin{equation}
    \varrho(p_{\rm mix})
    =p_{\rm mix}\ket{0}\!\bra{0}
    +(1-p_{\rm mix})\ket{1}\!\bra{1},
    \qquad p_{\rm mix}\in[0,1/2],
\end{equation}
whose $z$-aligned Bloch vector has length $1-2p_{\rm mix}$.  Increasing
$p_{\rm mix}$ removes purity, suppressing the maximum conditional-state
distinguishability and hence the record capacity.  In particular,
$p_{\rm mix}=1/2$ is maximally mixed
and cannot store a local record of $Z$, whereas our $p=1/2$ state is the pure
$x$-polarized state and fails only when its direction is blind to the
conditional rotation.

The relative-rotation description also applies to mixed witnesses. For the
conditional states $\varrho_{sj}=U_j^{(s)}\varrho_jU_j^{(s)\dagger}$ of a
qubit with initial Bloch vector $\mathbf r_j$, the Uhlmann root fidelity is
\begin{equation}
    B_j^2(\varrho_{0j},\varrho_{1j})
    =1-|\mathbf q_j\times\mathbf r_j|^2.
    \label{eq:discussion_mixed_fidelity}
\end{equation}
For independent mixed witnesses, the conditional fragment states also
remain products, so their root fidelity obeys
$B_{\Frag}=\prod_{j\in\Frag}B_j$.
For mixed states, however, this product alone does not determine the
fragment Holevo information.

For the diagonal preparation above, write $r=1-2p_{\rm mix}$, so that
$\mathbf r_j=-r\hat{\mathbf z}$. The purity of each initial witness is
\begin{equation}
\begin{aligned}
    \mathcal P\equiv\operatorname{Tr}(\varrho_j^2)
    &=\frac{1+r^2}{2}\\
    &=p_{\rm mix}^2+(1-p_{\rm mix})^2.
\end{aligned}
    \label{eq:discussion_witness_purity}
\end{equation}
Using the conditional Hamiltonian
coefficients $a_{xj}=-(\pi g_j/2)\cos\theta$ and
$a_{zj}=h_j-(\pi g_j/2)\sin\theta$ from
Eq.~\eqref{eq:methods_branch_coefficients}, we obtain
\begin{align}
    Q_j(t)&=q_{xj}^2+q_{yj}^2
      =\frac{a_{xj}^2}{\Omega_j^2}\sin^2(\Omega_jt),
      \label{eq:discussion_mixed_Q}\\
    B_j^2(t)&=1-r^2Q_j(t),
      \qquad \Omega_j^2=a_{xj}^2+a_{zj}^2.
      \label{eq:discussion_mixed_separation}
\end{align}
Here $Q_j(t)=0$ when $\Omega_j=0$. Equivalently, the branch separation is
$1-B_j^2=(2\mathcal P-1)Q_j$: purity sets its amplitude, while $Q_j$
contains the field and time dependence. Reducing purity at fixed initial
direction leaves the zeros and extrema in field and time unchanged as long
as $r>0$. With no field,
$Q_j(t)=\cos^2\theta\sin^2(\pi g_jt/2)$.
Thus the same conditional-rotation mechanism governs local record writing
throughout this mixed-state family, as illustrated in
Fig.~\ref{fig:supp_mixed_comparison}(a).

For equal system branch probabilities, the single-witness Holevo information
can also be written exactly:
\begin{equation}
    \chi_z^{(j)}(t)
    =H_2\!\left(\frac{1+r\sqrt{1-Q_j(t)}}{2}\right)
     -H_2\!\left(\frac{1+r}{2}\right).
    \label{eq:discussion_mixed_holevo}
\end{equation}
At $r=1$ this reduces to the pure-state expression; at $r=0$ it vanishes.
For independent witnesses, branch fidelities still multiply across a
fragment, but its Holevo information is not generally determined by that
product alone.

At a fixed Bloch-vector direction, reducing purity is a depolarizing
channel that commutes with both conditional unitaries. Thus, for independent
witnesses with fixed initial directions and otherwise identical dynamics,
reducing their purity cannot increase the $Z$-Holevo information of any
fragment at any time, in agreement with earlier mixed-environment results
\cite{ZwolakQuanZurek2009MixedEnvironment,ZwolakQuanZurek2010RedundantImprinting}.
This result concerns the fixed $Z$ measurement. It does not imply
monotonicity of Holevo information optimized over the system measurement
or of the SBS distance.
Matching the populations of our pure preparation and the diagonal mixed
preparation, $p=p_{\rm mix}$, is a different comparison. At matched
populations, the pure state has $\mathcal P=1$ and the diagonal state has
$\mathcal P=p^2+(1-p)^2$. Removing the
initial off-diagonal coherence projects the Bloch vector onto the $z$ axis,
generally changing both its length and direction. The resulting change in alignment can outweigh the loss
of purity, so the mixed preparation can carry more or less information
about $Z$ than its pure counterpart
[Fig.~\ref{fig:supp_mixed_comparison}(b)].

Record distinguishability and system decoherence also respond differently
to purity. For the local decoherence factor
$\gamma_j=\operatorname{Tr}(\varrho_jU_j^{(0)\dagger}U_j^{(1)})$,
\begin{equation}
    B_j^2-|\gamma_j|^2
    =|\mathbf q_j|^2\bigl(1-|\mathbf r_j|^2\bigr)\geq0.
\end{equation}
Shortening a Bloch vector can therefore weaken its conditional record
while increasing decoherence. An optimized SBS distance depends on both,
as well as on the choice of basis; it need not follow the local Holevo
information [Fig.~\ref{fig:supp_mixed_comparison}(b,c)]. Our small-environment
comparisons establish these distinctions directly. They do not reproduce
all SBS extrema of Ref.~\cite{MironowiczHorodeckiHorodecki2022SelfEvolution}
or determine mixed-state fragment thresholds and persistence in the larger
disordered environments.

Replacing the initial pure environment with a mixed state leaves the
system's dynamically stable $Z$ pointer states unchanged. This dynamical statement does not
require $Z$ to maximize every finite-time information measure. Indeed,
mixed witnesses can yield more Holevo information for a system measurement
in a rotated basis [Fig.~\ref{fig:supp_mixed_comparison}(d)], unlike the pure
conditional states considered in our main fragment calculations.
For identical independent witnesses and a fixed observed fragment, this
additional information is a finite-environment effect: it vanishes as the
environment grows. Its maximum over time can decay algebraically, as the
size comparison and bound in the Supplemental Material show.

The no-field, uniform-coupling model of
Ref.~\cite{TouilYanGirolamiEtAl2022Eavesdropping} is a special case of the
present overlap formula.  For its initially $z$-polarized witnesses and unit
interaction time, the local branch overlap is $B_j=|\sin\theta|$, so a fragment
obeys $B_{\Frag}^2=|\sin\theta|^{2|\Frag|}$.  This recovers the amplification mechanism of that model.
Nonidentical couplings and local self-evolution were already treated in
Ref.~\cite{ZwolakRiedelZurek2016SpinEnvironments}; here their role is
resolved for the C-INOT preparation geometry, with matched field
distributions and mean and lower-percentile fragment thresholds.

\subsection{Testing the system measurement basis}
\label{subsec:discussion_basis_test}

\begin{figure*}[tp]
    \centering
    \includegraphics[width=.70\textwidth]{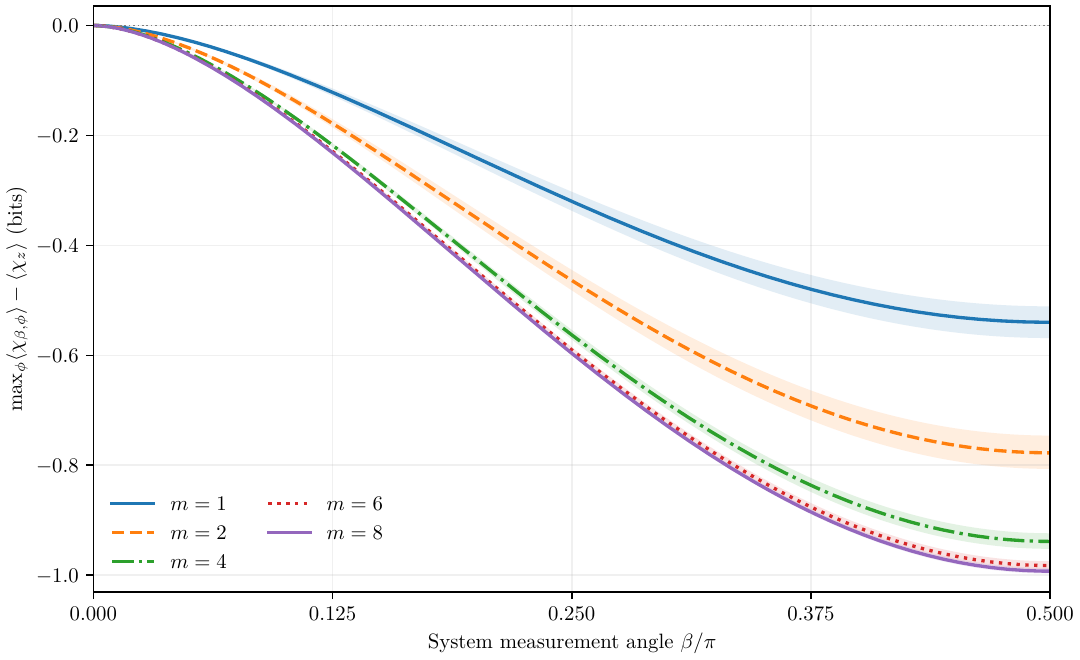}
    \caption{Testing the system measurement basis at $N_{\Env}=17$,
    $p=0.35$, $\theta=0.6$, $W=0.25$, and $t=60$, with no interactions
    between environment qubits.  Exact Holevo information relative to its
    $Z$-basis value, averaged over fragments within each of 24 realizations
    and then over realizations, before maximizing over azimuth at each
    polar angle $\beta$.  Curves show $m=1,2,4,6,8$; all sizes $1$--$8$
    are tested.  Shading gives 95\% intervals over realizations.
    Opposite measurement axes differ only by relabeling outcomes,
    allowing the range $0\leq\beta\leq\pi/2$.
    All 11\,160 sampled fragments attain their maximum at $Z$ on the
    tested grid.  Sampling and numerical checks are given in the
    Supplemental Material.}
    \label{fig:pointer_basis_test}
\end{figure*}

The controlled Hamiltonian is diagonal in the system's $Z$ projectors,
which select stable pointer states independently of environment purity
\cite{Zurek1981PointerBasis,DuruisseauTouilDeffner2023PointerStates,DoucetDeffner2024Classifying}.
Indeed, the branch decomposition in
Eq.~\eqref{eq:methods_branch_decomposition} gives
$U(t)=\sum_s P_s^{\Ssys}\otimes U_s(t)$, where
$P_s^{\Ssys}=|s\rangle\langle s|$ and $U_s(t)=e^{-iH^{(s)}t}$.
For any initial environment state $\rho_{\Env}$,
\begin{equation}
\begin{aligned}
    &U(t)\bigl[P_s^{\Ssys}\otimes\rho_{\Env}\bigr]U^\dagger(t)\\
    &\qquad=P_s^{\Ssys}\otimes U_s(t)\rho_{\Env}U_s^\dagger(t).
\end{aligned}
\label{eq:discussion_pointer_invariance}
\end{equation}
Thus an initially pure $Z$ eigenstate remains unchanged and uncorrelated
with the environment at every time, including for mixed environments.
This exact stability does not require the environment to retain a good record.
For the pure product environments considered here, measurement
in the $Z$ basis prepares pure conditional fragment states. Its conditional
entropy therefore vanishes, giving $\chi_z=H(\rho_{\Frag})$, the largest
possible Holevo information for any system measurement. This is the same
optimality argument used by Touil \emph{et al.}
\cite{TouilYanGirolamiEtAl2022Eavesdropping}.
Figure~\ref{fig:pointer_basis_test} confirms this property for the sampled
fragments and shows how the information decreases as the measurement axis
is rotated. Field-assisted recording in this setting consequently does not
require a change of the recorded system observable.

For mixed conditional states, the preceding pure-state argument
need not apply. A general bound still follows because the Holevo information
in any candidate system basis cannot exceed the mutual information:
\begin{equation}
    \max_{\mathbf n}\chi_{\mathbf n}-\chi_z
    \leq I(\Ssys{:}\Frag)-\chi_z
    =D_z.
    \label{eq:discussion_pointer_gain_bound}
\end{equation}
Any rotated-basis advantage is bounded by the pointer-basis
remainder and vanishes with it. For our pure conditional fragment states,
the stronger conclusion above already excludes such an advantage even
when $D_z$ is nonzero. Mironowicz \emph{et al.} instead define their
indicator basis by minimizing the distance to the closest SBS state.
The mixed-state examples in Fig.~\ref{fig:supp_mixed_comparison}
illustrate why these two optimizations must be distinguished; neither
a rotated information optimum nor a rotated closest-SBS basis changes
the invariant states of the controlled Hamiltonian.

\subsection{Why Holevo information and the pointer-basis remainder matter}
\label{subsec:discussion_discord}

Mutual information counts both the pointer-ensemble Holevo contribution and
correlations that are not independently readable records.  For the fixed
pointer basis, $\chi_z$ is the Holevo upper bound on information recoverable from a
fragment about the value of $\sigma_z^{\Ssys}$, while
\begin{equation}
    D_z(\Ssys{:}\Frag)=I(\Ssys{:}\Frag)-\chi_z(\Ssys{:}\Frag)
\end{equation}
is the part of the correlation not captured by those pointer records when the
system measurement is fixed to the pointer basis
\cite{HendersonVedral2001Correlations,ZwolakZurek2013DiscordAccessible}.  It is not discord minimized over all
system measurements.

The Holevo threshold tests the information in a fragment;
redundant recording additionally requires saturation at small fragment
sizes. High $\chi_z$ together with low
$D_z$ identifies a predominantly classical pointer record.  Low $D_z$ together
with low $\chi_z$ instead means that the fragment contains almost no
information, as occurs on the blind-alignment boundary.  Conversely,
a mutual-information plateau alone does not establish spectrum
broadcast structure or full conditional factorization
\cite{HorodeckiKorbiczHorodecki2015QuantumOrigins,LeOlayaCastro2018Objectivity,LeOlayaCastro2019StrongQD,Korbicz2021Roads,DoucetDeffner2025Compatibility}.
Together, high $\chi_z$ and low $D_z$ characterize predominantly
classical correlations in a finite fragment; they imply neither exact SBS nor an
approximate-SBS trace-distance bound.  Exact SBS additionally requires
the system to be decohered in the pointer basis, the conditional states of
different pointer values to have orthogonal support in each macrofraction, and
the macrofractions to be conditionally independent.  Strong QD---saturation of
both mutual information and the pointer-ensemble Holevo quantity---together with strong
independence, understood as full conditional factorization rather
than pairwise conditional independence alone, is equivalent to this structure
\cite{LeOlayaCastro2019StrongQD,FellerRousselFrerotDegiovanni2021Comment}.
Full conditional factorization holds exactly in the product model studied here;
the remaining obstructions are residual system coherence and nonzero
conditional branch overlaps.  Both
are products of single-witness quantities and can become exponentially small
without vanishing exactly at finite size.  Dynamical SBS monitoring therefore
tracks both decoherence and macrofraction distinguishability
\cite{MironowiczKorbiczHorodecki2017Monitoring}; rigorous finite-dimensional
SBS conditions are correspondingly stronger than a mutual-information plateau
\cite{AcevedoWehrKorbicz2024SBS}.

\subsection{Redundancy from imperfect local records}
\label{subsec:discussion_redundancy}

Finite-threshold redundancy does not require exact SBS or perfectly
distinguishable single-qubit states.  The mean-threshold index
$R_\delta=N_{\Env}/m_\delta$, and its quantile-based counterpart, summarize
fragment statistics.
Imperfect local witnesses can nevertheless combine into an accurate record
in a small fraction of the environment when branch separation
in that fragment is sufficiently large
\cite{ZwolakQuanZurek2010RedundantImprinting,RiedelZurek2010Everyday}.
In the noninteracting pure-product model,
\begin{equation}
    B_{\Frag}^{\rm typ}(t)\simeq
    e^{-|\Frag|\kappa_{\rm typ}(t)},
    \qquad
    \kappa_{\rm typ}(t)=-\mathbb E[\ln B_j(t)].
\end{equation}
For nonblind witnesses at generic times, $\kappa_{\rm typ}$ is positive.
The relevant measure of record formation is the fraction
$m_\delta/N_{\Env}$ needed for the Holevo information to approach
$H_Z(\Ssys)$.
A small positive exponent can leave the required fragment too large to
provide substantial redundancy.  Without self-fields, increasing $\Lambda$
increases the rate of record accumulation; self-fields can enhance or suppress it within
the observation window.  The inverse-rate estimate in
Eq.~\eqref{eq:results_threshold_estimate} captures this dependence, but
the actual mean and lower-quantile thresholds require the full distribution
of fragment overlaps
\cite{TuziemskiKorbicz2016BrownianSBS,ZwolakRiedelZurek2014Chernoff}.
The useful regime is therefore one in which accurate records are
available from small fragments, even though individual witnesses remain
imperfect.  This is a finite-information-threshold statement, not exact
objectivity at finite size
\cite{TouilYanGirolamiEtAl2022Eavesdropping}.

The product-law calculation in Fig.~\ref{fig:results_scaling} determines
the required fragment size from the local-overlap distribution.  Holding
that distribution fixed while enlarging the environment leaves the iid
threshold unchanged, so the same record occupies a smaller fraction of the
environment and the redundancy index grows.  The state-vector calculations
independently check the predicted fragment information at the sampled
environment sizes and the finite-fragment sampling procedure.  The product
model directly predicts this size dependence.

\subsection{Limitations}
\label{subsec:discussion_limitations}

The principal limitation is that the macrofraction calculations use pure,
initially uncorrelated environment qubits.  Although the local mixed-state
fidelity retains the relative-rotation form given above, the information in mixed conditional fragment states cannot in general
be determined from a single overlap
\cite{ZwolakQuanZurek2009MixedEnvironment}.  The local mixed-state comparisons in Sec.~\ref{subsec:discussion_mironowicz}
do not determine how mixed preparations change the fragment thresholds
and persistence in the larger disordered environments.

We also set $H_{\Env\Env}=0$.  This isolates the writing of the record but excludes its
subsequent transport, scrambling, or dynamical protection by interactions
within the environment.  Those processes require different diagnostics and
belong to a separate study.

Record quality can vary across fragments and in time
\cite{RiedelZurekZwolak2012RiseFall}, so we assess persistence directly using
the minimum of the half-environment 10th-percentile Holevo information over
the observation window.  A realization passes when this minimum meets
the record threshold: useful records then remain available at every sampled
time, rather than only on average.  The fraction of realizations that pass
measures how consistently this persistence occurs across coupling and field
realizations.  Paired samples support comparisons between parameter choices,
and the uncertainty analysis in the Supplemental Material tests sensitivity
to the sampled realizations and fragments.  This provides evidence for
persistent records over the specified observation windows and for their
robustness within the coupling and field distributions studied here.
Near the threshold, the reported intervals and passing counts indicate the
strength of that evidence.

%% file: sections/conclusions.tex
\section{Conclusions}
\label{sec:conclusions}

Environment-state alignment controls record writing in the C-INOT model.
The local equivalence of each gate to CNOT preserves its entangling power,
and the controlled Hamiltonian fixes the system's $Z$ pointer observable.
The preparation and relative conditional rotation determine how much branch
information each witness acquires.  Without self-fields, one invariant
$\Lambda$ determines all overlap-derived diagnostics and identifies exact
blind preparations.  Simulations with different initial environment states
and interaction axes, chosen to have the same $\Lambda$, reproduce the
same curves for fragment information.
Increasing the interaction angle can therefore improve or suppress
recording, depending on the initial preparation: gate imperfection alone
does not determine record quality. The relative-rotation description also
applies to other controlled-unitary interactions with environment qubits,
although their dynamics need not reduce to a single constant $\Lambda$.
For our pure product environments, $Z$ also maximizes fragment Holevo
information, so field-assisted recording requires no change of system basis.

Touil \emph{et al.} established amplification of imperfect
records in the C-INOT family, and Zwolak, Riedel, and Zurek treated
preparation alignment and self-evolution in nonuniform spin environments
\cite{TouilYanGirolamiEtAl2022Eavesdropping,ZwolakRiedelZurek2016SpinEnvironments}.
Our contribution resolves the preparation, interaction-angle, and field
dependence of C-INOT records and compares uniform local field strengths with
strengths drawn from a Gaussian distribution at equal root-mean-square
strength. It quantifies how field disorder changes record quality and how
reliability requirements increase the necessary fragment size.

Local fields can create local branch separation in an initially blind preparation,
restore recording, or delay it beyond the observation window. Their effect
requires noncommuting field and recording axes; at the commuting endpoint,
the local fields leave record quality unchanged. Strong fields
suppress transverse motion, while the residual response at the aligned
preparation occurs on the dispersive timescale $O(|h_j|/g_j^2)$.
At the initially blind point $p=1/2$, $\theta=0$, and equal rms strength,
zero-mean Gaussian fields broaden both the weak-field improvement
and the strong-field suppression.  Uniform fields perform better near the
onset, whereas Gaussian fields retain more information at strong fields.
Their different revival patterns also affect persistence, which we test
through the minimum record quality over the sampled window and the fraction
of realizations meeting the criterion.
A modest spread around a nonzero mean already suppresses the
intermediate-field dips of the uniform profile. Effective recording occurs
across broad regions of the sampled preparations and fields, rather than
requiring a finely tuned initial state.

Low pointer-basis remainder $D_z$ alone does not establish a good record:
it also occurs when fragments carry almost no information. Predominantly
classical redundant records require high Holevo information already in
small fragments, saturation as fragment size increases, and a small
remainder over that range. For the pure conditional states studied here,
the overlap also fixes the optimal branch-discrimination error, connecting
the Holevo criterion to record readout without identifying Holevo and
accessible information. These finite-threshold diagnostics do not by
themselves establish exact spectrum broadcast structure.

For independent pure witnesses, the fragment overlap is an exact product.
Its distribution determines both mean and lower-quantile Holevo thresholds,
quantifying the extra fragment size needed when record quality varies
among environment qubits.  For independent, identically distributed witnesses with a
fixed local-overlap distribution, the fragment size needed to meet a chosen
information threshold stays fixed as the environment grows.  The same
record therefore occupies a smaller fraction of the environment, and the
corresponding redundancy $N_{\Env}/m_\delta$ grows linearly with its size.
The state-vector calculations independently validate selected predictions
for fragment information and the sampling procedure.
Requiring most fragments to pass the information threshold needs larger
fragments than a threshold based on the mean alone, but preserves this
qualitative size dependence. The finite environments studied here exhibit
modest redundancy; the fixed-threshold result explains how it can grow.
High mean information also need not imply persistence. For suitable fields,
our lower-quantile criterion remains satisfied at every sampled time in
the late observation window.

The relative-rotation explanation extends to the diagonal mixed
witnesses considered by Mironowicz \emph{et al.}
\cite{MironowiczHorodeckiHorodecki2022SelfEvolution}.
Our analytical expressions and small-environment calculations verify
that, at fixed initial Bloch-vector direction, reducing
purity weakens local branch separation without changing its zeros
or the positions of its extrema in field and time, provided the Bloch
vector remains nonzero (Sec.~\ref{subsec:discussion_mironowicz} and
Fig.~\ref{fig:supp_mixed_comparison}).
For independent witnesses at fixed initial directions and otherwise identical
dynamics, reducing purity cannot increase the $Z$-Holevo information of any
fragment. Removing coherence at fixed populations generally changes the initial direction,
however, and the resulting change in alignment can improve or suppress
recording. The Hamiltonian preserves the same system $Z$ pointer states at
every purity, even when a rotated measurement yields more finite-time
Holevo information. For identical independent witnesses and a fixed observed
fragment, this advantage vanishes as the environment grows.
The conditional fidelity of
a mixed fragment remains the product of its local fidelities, extending
this mechanism beyond individual witnesses. Larger-fragment calculations
would quantify how purity changes redundancy and persistence.
With interactions within the environment, those conditional states would
also track how transport and scrambling redistribute or erase the records.

%% file: sections/acknowledgments.tex
\begin{acknowledgments}
A.L. acknowledges funding from the European Union's Horizon Europe research
and innovation programme under the Marie Sk{\l}odowska-Curie grant agreement
No.~101210678 (QDGPUS).  Views and opinions expressed are
however those of the author(s) only and do not necessarily reflect those of
the European Union or the European Research Executive Agency (REA).  Neither
the European Union nor the granting authority can be held responsible for
them.
\end{acknowledgments}

%% file: sections/data_availability.tex
\section*{Data and Code Availability}

Simulation software, batch configurations, exact-analysis scripts, and
processed figure data are available in \texttt{pySL\_qubits}
\cite{Lasek2026pySLQubits}.  The included
\href{https://github.com/ALasek/pySL_qubits-public/tree/paper-a-v1.2.0/paper_a}{reproduction package}
redraws all plots.  Full raw simulation outputs are available
from the authors upon reasonable request.

%% file: sections/supplement.tex
\clearpage
\onecolumngrid
\setcounter{figure}{0}
\renewcommand{\thefigure}{S\arabic{figure}}
\renewcommand{\theHfigure}{S\arabic{figure}}

\section*{Supplemental Material}

\subsection*{Different preparations with the same \texorpdfstring{$\Lambda$}{Lambda}}

\begin{figure}[h!]
    \centering
    \includegraphics[width=\textwidth]{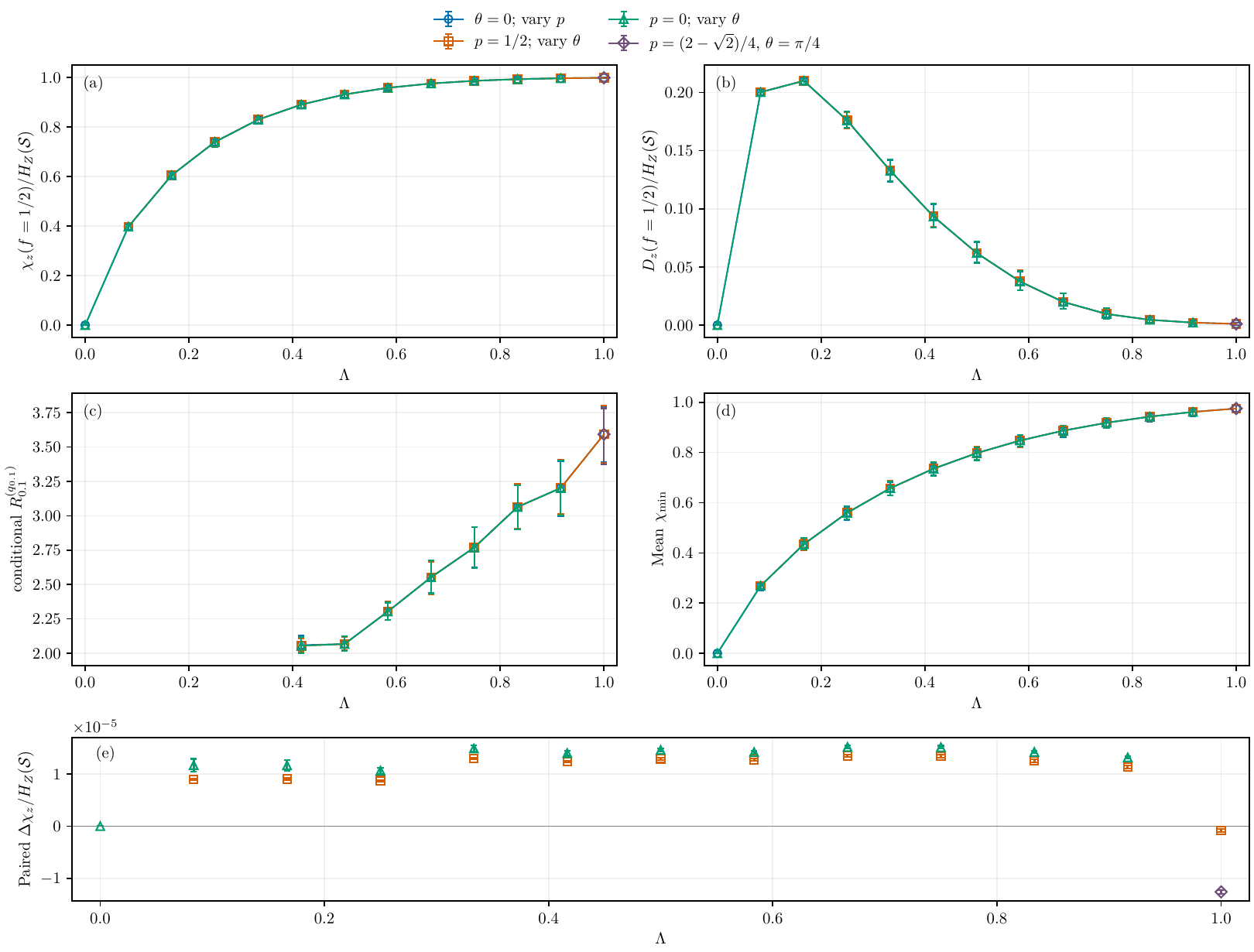}
    \caption{Comparison of different preparations and interaction angles
    at fixed $\Lambda$, with $N_{\Env}=16$, no local fields, and no
    environment--environment interaction.  The scan varies $p$ at $\theta=0$
    or varies $\theta$ at $p=1/2$ and $p=0$.  It contains 38 parameter points
    at 13 values of $\Lambda$, including an additional check at $\Lambda=1$.
    Panels (a,b) show Holevo information and pointer-basis
    remainder averaged over half-environment fragments.  Panel (c) uses the smallest
    fragment size whose 10th-percentile Holevo information meets the $90\%$
    threshold to define $R_{0.1}^{(q_{0.1})}$.  Panel (d) uses the minimum
    over time of that percentile at half-environment size,
    Eq.~\eqref{eq:methods_holevo_min}.  Information quantities are divided
    by $H_Z(\Ssys)$.  The window is $t\in[20,40]$; panels (a--c) average
    the time medians of the 24 realizations, and panel (d) averages their
    time minima.  The redundancy medians include only times with a
    threshold crossing.  Panel (e) shows differences in the time-median
    Holevo information averaged over fragments, relative to the $\theta=0$ preparation at the same
    $\Lambda$, calculated within paired realizations before averaging.
    Error bars are
    95\% confidence intervals; panel (e) resamples paired realizations.}
    \label{fig:supp_matched_lambda_collapse}
\end{figure}

This check tests the prediction that $\Lambda$ alone determines the
information measures in the no-field model.  The parameter points at
each $\Lambda$ use the same 24 coupling realizations and the same sampled
fragments (up to 64 per size).  We compare these points before averaging
them in the main text.
The largest differences between their central estimates are
$1.52\times10^{-5}$ for half-environment Holevo information,
$2.04\times10^{-5}$ for the pointer-basis remainder,
$4.58\times10^{-3}$ for $R_{0.1}^{(q_{0.1})}$, and
$1.55\times10^{-5}$ for the mean time minimum in panel (d).
These small differences quantify the numerical agreement with the exact
prediction.  The intervals quantify uncertainty in the means in panels
(a--d) and in the mean paired differences in panel (e).

\clearpage
\subsection*{Checks with larger environments}

\begin{figure}[htp]
    \centering
    \includegraphics[width=.90\textwidth]{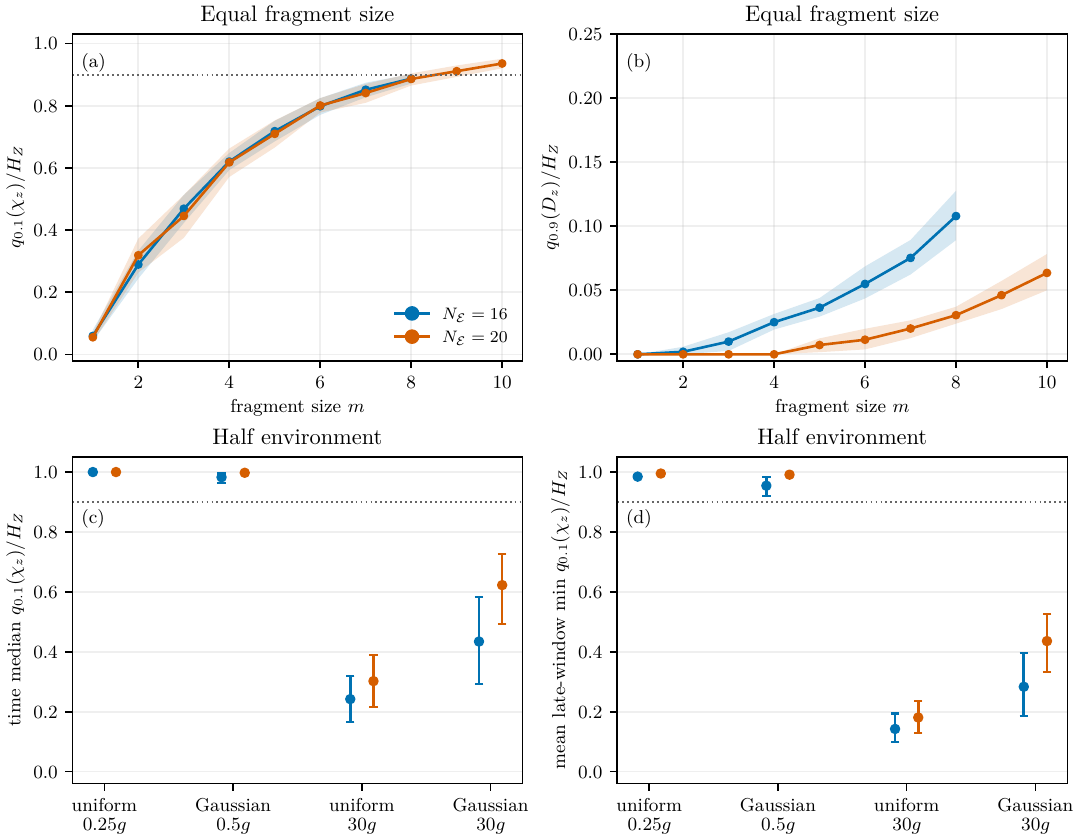}
    \caption{Numerical checks at $N_{\Env}=20$ against the
    matching $N_{\Env}=16$ cases.  The top row compares equal fragment
    sizes; the bottom row compares equal fractions $f=1/2$.
    Both sizes use the same eight random seeds, with up to 64 fragments
    per size at $N_{\Env}=16$ and 32 at $N_{\Env}=20$.
    All fragments are used when fewer exist.  Panels (a) and (b) show the 10th-percentile Holevo
    information and 90th-percentile pointer-basis remainder for the no-field
    $p=1/2$, $\theta=\pi/4$, $\Lambda=1/2$ case, summarized over
    $t\in[20,40]$.  Panels (c) and (d) use the half-environment
    10th-percentile Holevo information for four field settings at the
    initially blind preparation $p=1/2$, $\theta=0$, over $t\in[30,60]$.
    All quantiles are calculated within each realization.  Panels (a--c)
    average time medians across realizations, and panel (d) averages time
    minima.  Information quantities are divided by $H_Z(\Ssys)$.
    Bands and error bars are 95\% confidence intervals over the eight realizations.}
    \label{fig:supp_higher_n_validation}
\end{figure}

These runs check both accumulation of information in larger fragments
and the field response at a larger environment size.  They confirm the
selected cases of field-induced record formation and strong-field delay.
The size dependence predicted by the product-overlap model is tested
separately in Fig.~\ref{fig:results_scaling}.

A separate no-field comparison uses the same eight realization seeds at
$N_{\Env}=16,20,24$ and the fixed time $t=36$, for $p=1/2$ and
$\theta=\pi/4$.  Unlike the time-window summaries in
Fig.~\ref{fig:supp_higher_n_validation}, it tests record quality at one time.
The $N_{\Env}=16$ runs use up to 64 fragments per size; the larger runs use 32.
\begin{center}
\begin{tabular}{cccc}
\toprule
$N_{\Env}$ & Mean half-environment $q_{0.1}[\chi_z]/H_Z$ & 95\% interval & Above $0.9$ at $t=36$\\
\midrule
16 & $0.890$ & $[0.869,0.909]$ & $2/8$\\
20 & $0.942$ & $[0.928,0.954]$ & $8/8$\\
24 & $0.967$ & $[0.960,0.973]$ & $8/8$\\
\bottomrule
\end{tabular}
\end{center}
At $N_{\Env}=24$, the quantile-based redundancy is defined in all eight
realizations, with mean $2.75$ and a 95\% interval $[2.67,2.88]$.

\clearpage
\subsection*{Sensitivity to the fragment percentile}
\begin{figure}[h!]
    \centering
    \includegraphics[width=\textwidth]{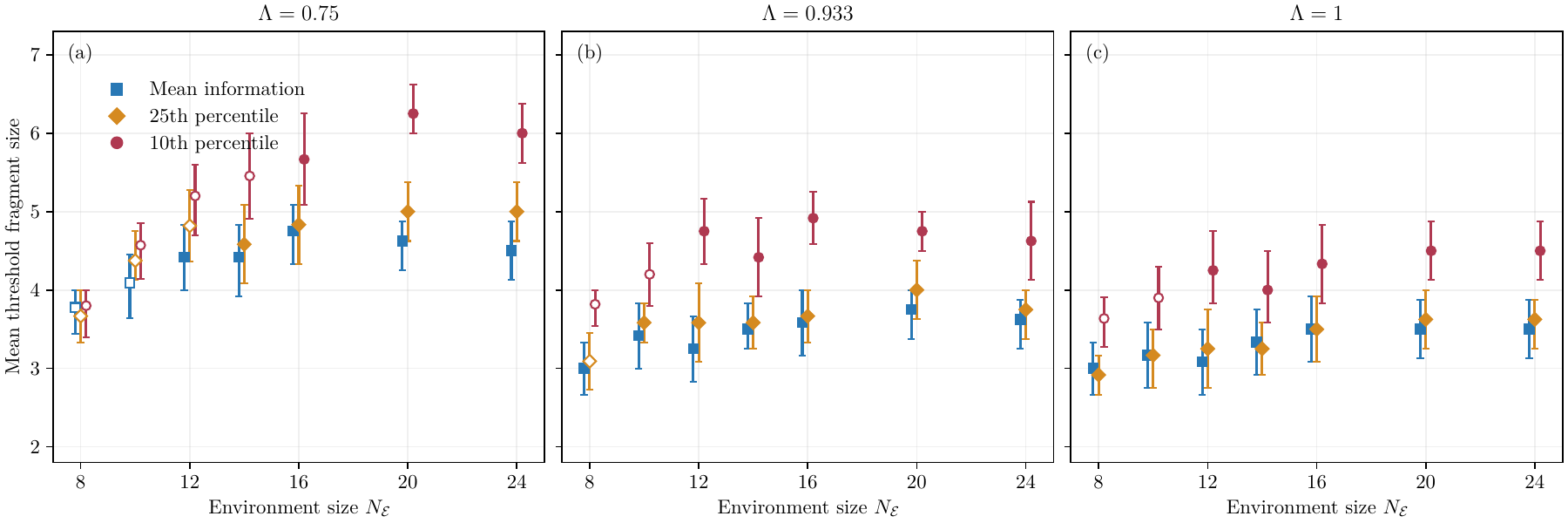}
    \caption{Mean required fragment size at $t=36$ for three ways of
    applying the $90\%$ information threshold: mean Holevo information,
    its 25th percentile, and its 10th percentile.  Calculations use the same
    couplings and fragment selections as Fig.~\ref{fig:results_scaling}(b).
    Sizes are found within each realization before averaging.
    Open symbols exclude realizations without a crossing among
    $m\leq N_{\Env}/2$; filled symbols include all realizations.
    Bars are 95\% confidence intervals obtained by resampling realizations
    with their fragments held fixed.}
    \label{fig:supp_percentile_comparison}
\end{figure}
The comparison covers all 57 parameter cases used in
Fig.~\ref{fig:results_scaling}(b).  At $N_{\Env}=24$, all realizations cross
under all three criteria.  The 10th-percentile criterion increases the mean
required size by $28$--$33\%$ relative to the mean-information criterion;
the 25th percentile increases it by $4$--$11\%$.
In the 12 cases retaining individual information samples, all three
thresholds also agree with the state-vector calculation.

\subsection*{Sampling for the required fragment size}

Figure~\ref{fig:results_scaling}(b) uses $t=36$ and the same couplings and
fragments for exact and state-vector calculations.  For each realization,
we find the first crossing of the 10th-percentile Holevo threshold among
$m\leq N_{\Env}/2$ and retain that fragment size.  At $N_{\Env}=8,12,16$,
we average these sizes over six, four, and five equivalent
preparation--angle pairs for $\Lambda=0.75,0.933,1$, respectively, keeping
the realization seeds paired.  Other sizes use one preparation--angle pair.
We then average over realizations with a crossing.

The $N_{\Env}=8$--$16$ cases use 12 realizations and up to 32 fragments per
size; $N_{\Env}=20,24$ use eight realizations and up to 64 fragments per
size.  All three values of $\Lambda$ are sampled at every size.
For $\Lambda=(0.75,0.933,1)$, the numbers of realizations with crossings are
$(5,11,11)$ at $N_{\Env}=8$, $(7,10,10)$ at $10$, $(10,12,12)$ at $12$,
and $(11,12,12)$ at $14$.  All realizations cross at the remaining sizes.
Exact and state-vector calculations give identical thresholds in all 57
parameter cases.  Their individual 10th-percentile Holevo values differ
by at most $7.84\times10^{-3}$ bits.
Intervals use 2000 bootstrap resamples of realizations with a threshold
crossing, keeping their fragments and matched preparations together.
They quantify uncertainty in the mean over those realizations.

\subsection*{System measurement basis}
Figure~\ref{fig:pointer_basis_test} uses exact conditional states
in the two-dimensional span of the fragment branches.  Each of 24
realizations uses all 17 single-qubit fragments and 64 sampled fragments
at each size $m=2,\ldots,8$.  The grid has 181 uniformly spaced polar
angles in $[0,\pi/2]$, 24 additional logarithmically spaced angles near
$Z$ starting at $10^{-4}$ radians, and 80 azimuths in $[0,2\pi)$.
Intervals use 2000 resamples of realizations, holding fragments fixed
and again finding the azimuth that maximizes the mean Holevo information
at each polar angle.  Computing the eigenvalues from the trace and
determinant agrees with direct matrix diagonalization within
$4.22\times10^{-15}$ bits in
representative checks across all sizes.  The original
240 state-vector scans at $m=4,6$ agreed with their exact reconstruction
within $5.58\times10^{-5}$ bits and also selected $Z$.

\clearpage
\subsection*{Field response at \texorpdfstring{$\theta=0$ and $\theta=\pi/2$}{theta = 0 and theta = pi/2}}

\begin{figure}[h!]
    \centering
    \includegraphics[width=\textwidth]{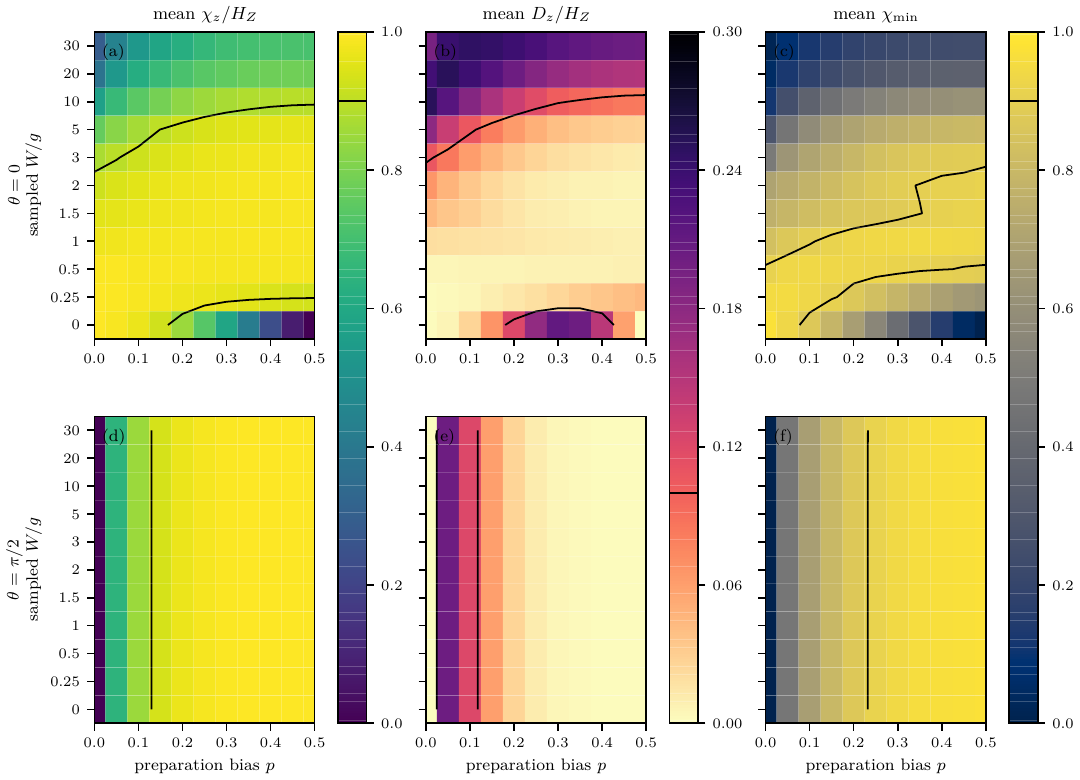}
    \caption{Field response evaluated with exact product overlaps and no
    information cutoff.  Rows show $\theta=0$ and
    $\theta=\pi/2$; columns show normalized mean half-environment Holevo
    information, normalized mean pointer-basis remainder, and the
    time minimum of the half-environment 10th-percentile Holevo
    information.  Each $11\times11$ map samples $p\in[0,0.5]$ and
    $W/g\in\{0,0.25,0.5,1,1.5,2,3,5,10,20,30\}$; these sampled values are
    shown with equal spacing, despite the unequal intervals between field
    strengths.  The calculations use $N_{\Env}=16$, $g=0.1$, 12
    realizations per point, and no environment--environment
    interaction.  We use the 21 sampled times in $t\in[20,40]$.
    The first two columns average time medians across realizations;
    the third averages their time minima.  All information quantities
    are divided by $H_Z(\Ssys)$.  Black contours mark levels $0.9$, $0.1$,
    and $0.9$ in the three columns, interpolated between sampled grid
    points as visual guides.}
    \label{fig:supp_stability_endpoints}
\end{figure}

The two angles test whether the field and conditional interaction
commute.  At $\theta=\pi/2$, both act along $Z$.  The self-field then
cancels from the relative branch evolution, so changing $W$ leaves the
record unchanged.  At $\theta=0$, the interaction acts along $X$ and
the $Z$ field changes the relative evolution.  This comparison isolates
the role of the interaction axis in the field response.

\clearpage
\subsection*{Uniform and Gaussian fields with a commuting interaction}

\phantomsection\label{sec:supp_commuting_control}

At $p=1/2$ and $\theta=\pi/2$, the relative branch evolution is independent
of the local $Z$ field.  We check this numerically using uniform and
zero-mean Gaussian profiles at $h_{\rm rms}/g=0,0.5,2,10,30$, with
$N_{\Env}=16$, 24 paired realizations, and up to 64 fragments per size.
Over $t\in[30,60]$, the largest variations among these nine cases are:
\begin{center}
\begin{tabular}{lc}
\toprule
Half-environment statistic & Maximum variation\\
\midrule
Mean Holevo information, time median & $2.03\times10^{-5}$\\
Mean pointer-basis remainder, time median & $2.39\times10^{-5}$\\
10th-percentile Holevo information, time minimum & $2.0\times10^{-4}$\\
\bottomrule
\end{tabular}
\end{center}
All information is divided by $H_Z(\Ssys)$; each time statistic is computed
within a realization and then averaged.  These small numerical residuals
check the field independence illustrated in
Fig.~\ref{fig:supp_stability_endpoints}.

\subsection*{Additional checks at weak fields}
Eight additional cases at the initially blind preparation
$p=1/2$, $\theta=0$ use uniform and Gaussian fields with
$h_{\rm rms}/g=0.05,0.10,0.15,0.20$.
They use the same 24 random seeds and fragment sampling as the field comparison,
with the information cutoff disabled.  The exact calculation
reproduces every saved fragment selection.  For $m\leq8$, its largest
difference from the simulated Holevo information averaged over fragments is
$2.46\times10^{-5}$ bits.  No realization changes its classification
as persistent or nonpersistent over the sampled times.
The two-stage bootstrap described below uses 1000 draws for these cases.
Its mean differs from the original estimate by at most $0.00105$;
its 95\% interval widens by at most $0.00470$ compared with resampling
realizations alone.
At $0.20g$, these intervals for the mean minimum are
$[0.929,0.947]$ for uniform fields and $[0.730,0.842]$ for Gaussian fields.

\clearpage
\subsection*{Gaussian fields with a nonzero mean}

To separate field strength from the spread of local fields, we interpolate
between uniform and zero-mean Gaussian profiles at fixed rms strength:
\begin{equation}
    h_j=h_{\rm rms}\left(\sqrt{1-\eta^2}+\eta\xi_j\right),
    \qquad \xi_j\overset{\rm iid}{\sim}\mathcal N(0,1).
    \label{eq:supp_field_mean_width}
\end{equation}
Thus $h_0=h_{\rm rms}\sqrt{1-\eta^2}$ and $W=\eta h_{\rm rms}$ in
Eq.~\eqref{eq:methods_coupling_distributions}.  The parameter $\eta$ sets
how much of the rms strength comes from disorder: $\eta=0$ gives a uniform
field, $\eta=1$ a zero-mean Gaussian, and intermediate values a Gaussian
with nonzero mean.  The rms strength refers to the distribution;
individual realizations are not rescaled.

\begin{figure}[h!]
    \centering
    \includegraphics[width=\textwidth]{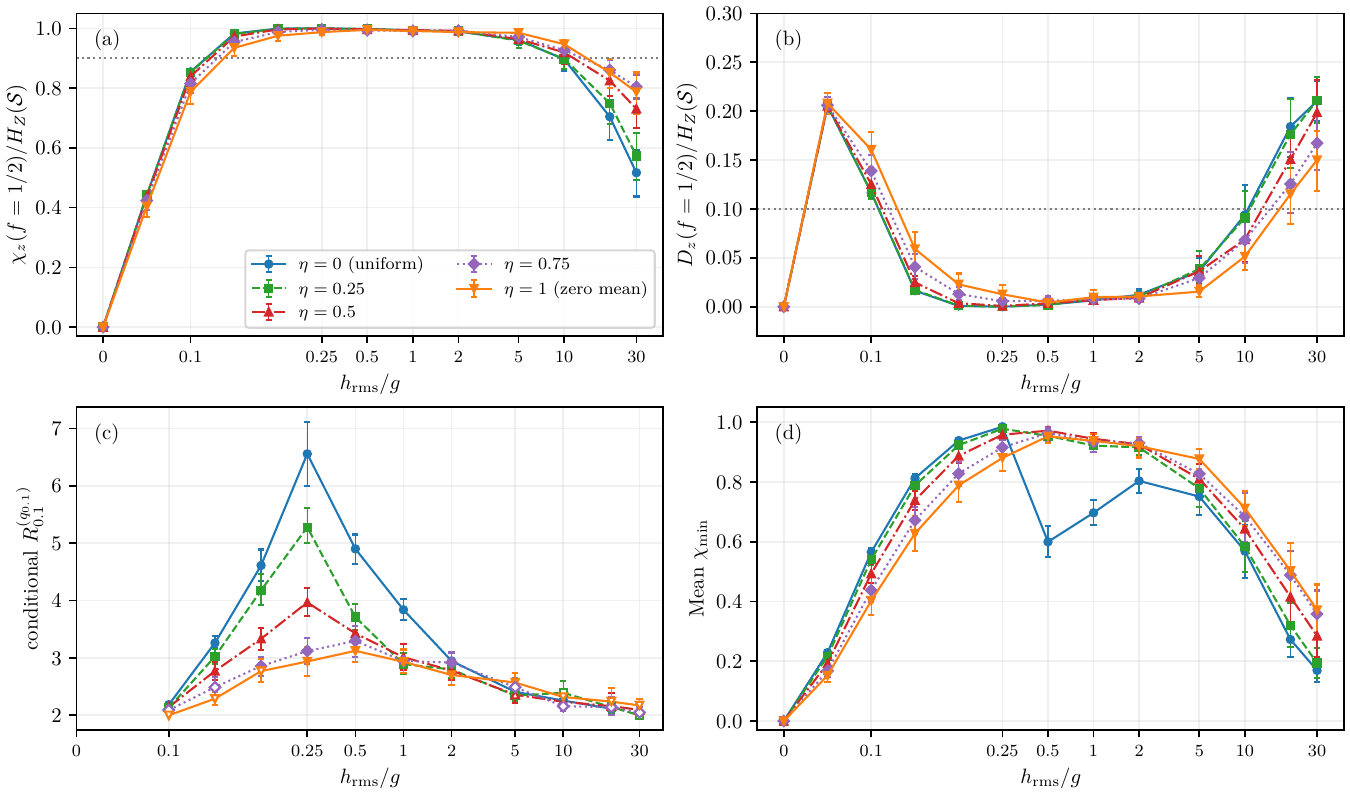}
    \caption{Field mean and width at fixed rms strength, with
    $p=1/2$, $\theta=0$, $N_{\Env}=16$, and $g=0.1$.
    Exact product overlaps use 24 paired realizations and up to 64
    fragments per size, with the same coupling draws, standardized field
    draws, and fragment selections across cases.
    Panels (a,b) show $\chi_z$ and $D_z$ averaged over half-environment fragments,
    summarized by time medians within each realization and then averaged.
    Panel (c) shows $R_{0.1}^{(q_{0.1})}$.  Within each realization, we take
    its median over times with a threshold crossing, then average these
    medians across realizations, as in Fig.~\ref{fig:results_self_evolution_map}.
    Open symbols indicate that fewer than 24 realizations have a crossing;
    no point is shown when none does.
    Panel (d) averages the time minima of the half-environment 10th percentile,
    Eq.~\eqref{eq:methods_holevo_min}.
    All time summaries use integer times in $[30,60]$.
    Bars are 95\% intervals from 2000 resamples of whole realizations,
    holding their sampled fragments fixed.
    The quantities and field axis match Fig.~\ref{fig:results_self_evolution_map}.}
    \label{fig:supp_field_mean_width}
\end{figure}

The nonzero-mean profiles show that improved persistence does not require
zero mean.  A modest spread already reduces the intermediate-field dips
of the uniform profile while retaining high Holevo information.
Near the onset of recording, the same spread instead lowers record quality.
At strong fields, broader distributions retain more information without
generally restoring persistence.  These results support an explanation
based on the distribution of local dynamics, including suppression of
revivals.  At this preparation, changing $h_j$ to $-h_j$ leaves $B_j$
unchanged by Eq.~\eqref{eq:methods_relative_components}; cancellation
between positive and negative fields cannot account for the difference.

As a numerical check, 18 cases with $h_{\rm rms}/g\leq0.2$ were also
evolved as full state vectors with $\Delta t=0.002$.  The sampled 10th-percentile Holevo values
agree with the exact calculation to within $2.90\times10^{-5}$ bits,
and all persistence classifications agree.  At $h_{\rm rms}/g=0.5$,
changing $\eta$ from $0$ to $0.25$ increases the mean minimum by $0.355$,
with a paired 95\% bootstrap interval $[0.307,0.402]$.  This interval uses
10\,000 resamples of whole realizations with fragments held fixed.

\clearpage
\input{sections/supplement_theta_fields}
\clearpage
\input{sections/supplement_mixed_states}
\clearpage

\subsection*{Numerical details}

All times use the unit $\hbar/E_0$ defined in the main text.
Information is measured in bits unless normalized by $H_Z(\Ssys)$.

\textit{Integrator checks.}  Norm and reduced-density-matrix checks are
available for the final realization of each $N_{\Env}=20$ run.
For these realizations, the state norm deviates from unity by at most
$6.04\times10^{-5}$, and the trace of a reduced density matrix by at most
$1.2\times10^{-4}$.  Hermiticity is monitored separately.
We also repeat four paired realizations of the no-field
$\Lambda=1/2$ case and of the Gaussian-field cases at $0.5g$ and $30g$ with
$\Delta t=0.001$.  After halving the time step, we compare the normalized
half-environment $q_{0.1}(\chi_z)$ and mean $D_z$ at each saved time and
realization in the late windows.  Their largest change is $2.3\times10^{-3}$;
the averages over realizations change by at most $6.21\times10^{-4}$.
The record classifications and the smallest fragment sizes meeting the
Holevo threshold remain unchanged at every sampled time.
These comparisons measure sensitivity to the time step; they do not bound
all discretization and rounding errors.

\phantomsection\label{par:supp_information_cutoff}
\textit{Exact reference and information cutoff.}  The original state-vector
analysis sets $I$ and $\chi_z$ to zero below $0.01$ nats ($0.01442695$ bits).
It then forms $\max(0,I-\chi_z)$ from those values and applies the same
cutoff.  The revised information plots use exact product overlaps without
this cutoff, with $I=\chi_z+D_z$.
We compared all 512 original parameter cases with this exact reference.
Every realization keeps the same persistence classification.  At $t=36$,
the checked fragment thresholds also remain unchanged: these tests average
10th percentiles calculated within realizations and compare their mean
with $0.9H_Z(\Ssys)$.  This earlier cutoff check uses a threshold of the
averaged quantile; the separate check for Fig.~\ref{fig:results_scaling}(b)
above applies thresholds within each realization.
The largest difference in the mean time minimum is $4.59\times10^{-3}$.
For fragments up to half the environment, the largest difference in a
10th percentile is $0.0219$ bits.  Integration, rounding, and the cutoff
all contribute to these observed differences.
Saved fragment selections are reproduced in all 74 cases that retain them.
For the other 438 cases, the selections are reconstructed from the original
random seeds and sampling algorithm.  The comparison uses every evaluated
fragment size and time.

\textit{Fragment sampling.}  At each size $m$, the fragments are uniformly
sampled unordered subsets of the environment sites.  The same fragment subsets
are used at all times and across matched parameter sweeps.  Subsets of
different sizes are drawn independently, so a size-$m$ fragment need not
contain a sampled size-$(m-1)$ fragment.  The runs with larger environments
reuse the first eight realization seeds of the $N_{\Env}=16$ runs.

\textit{Confidence intervals.}  The plotted intervals resample whole
realizations while keeping their sampled fragments fixed.
The interaction-angle sweeps in Fig.~\ref{fig:supp_theta_fields}
use 2000 resamples at each of 121 angles.
To check sensitivity to fragment selection as well, we use a two-stage
(nested) bootstrap with 1000 draws per parameter case for the original
512 exact calculations.  Each draw first samples realizations with replacement.
For each selected realization, it then samples fragments with replacement
from that realization's saved set.  Repeated selections of a realization
receive independent fragment draws.  Each selected fragment contributes
its complete time series, preserving correlations between times.
We calculate the 10th percentile across fragments at each time, take its
minimum over the window, and average those minima across the selected
realizations.

Across these cases, the nested bootstrap mean differs from the original
central estimate by at most $0.0262$, and its 95\% percentile interval
widens by at most $0.0304$ relative to resampling realizations alone.  The
nested mean lies on the opposite side of $0.9$ from the original estimate
in eight cases.  This shows sensitivity near the threshold; the figures
keep the original estimates.  In the $\theta=\pi/4$ map, 85 points have
original means at least $0.9$, and 70 have a lower endpoint of the nested
interval above $0.9$.
Each interval applies to one parameter point, not to the map as a whole.
For this minimum statistic, nominal 95\% bootstrap intervals have no
guaranteed 95\% coverage.  Selected results at $p=1/2$, $\theta=0$ are:
\begin{center}
\begin{tabular}{lcccc}
Profile & $h_{\rm rms}/g$ & Mean minimum & Nested 95\% interval & Passing realizations\\
\hline
Uniform & $0.25$ & $0.985$ & $[0.981,0.988]$ & $24/24$\\
Gaussian & $0.5$ & $0.953$ & $[0.929,0.971]$ & $22/24$\\
Gaussian & $1$ & $0.938$ & $[0.908,0.960]$ & $19/24$\\
Gaussian & $2$ & $0.920$ & $[0.880,0.946]$ & $17/24$
\end{tabular}
\end{center}
Here, a passing realization meets the $90\%$ threshold at every sampled
time, using its half-environment fragment 10th percentile.

\textit{Late-time window.}  Without self-fields, averaging the local
oscillation $\sin^2(\pi g_jt/2)$ over the Gaussian coupling distribution gives
$\mathbb E[\sin^2(\pi g_jt/2)]=\tfrac12\bigl(1-e^{-\pi^2\sigma_g^2t^2/2}\bigr)$,
with decay time $\sqrt2/(\pi\sigma_g)\simeq4.5$ for $\sigma_g=0.1$.
The exponential term is below $10^{-8}$ at $t=20$, motivating the start of
the no-field averaging window.  Individual realizations still oscillate;
averaging over couplings does not make them stationary.
This estimate does not generally apply when self-fields are present.
In particular, strong fields can delay record writing into or beyond the
chosen window.
Except for the persistence statistic, time summaries use medians within
each realization.  Redundancy medians include only times with a threshold
crossing.  All quantities in a comparison use the same window.

%% file: sections/supplement_theta_fields.tex
\subsection*{Interaction-angle dependence with local fields}

\begin{figure}[h!]
    \centering
    \includegraphics[width=\textwidth]{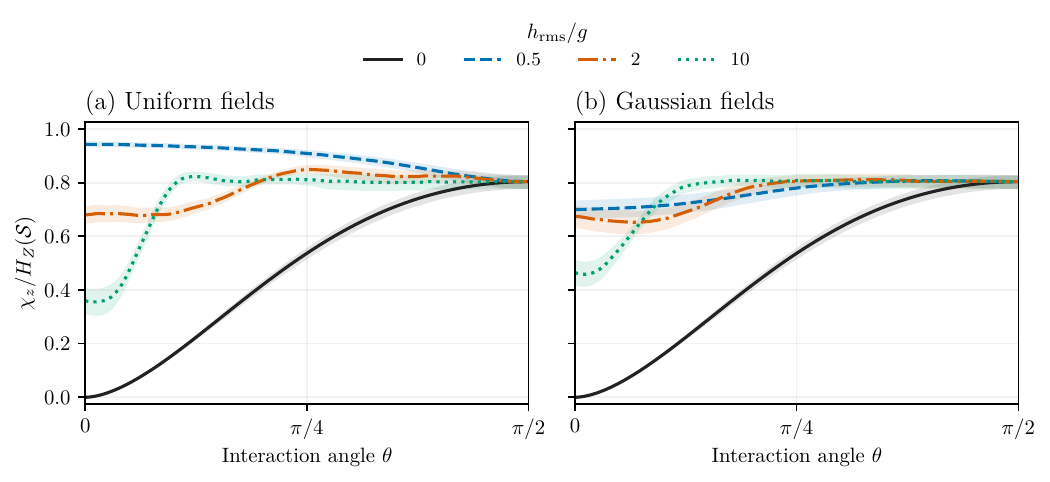}
    \caption{Fragment Holevo information versus interaction angle
    for (a) uniform and (b) zero-mean Gaussian fields, at matched rms
    strengths $h_{\rm rms}/g$ and pure preparation $p=1/2$.
    Both panels use $N_{\Env}=16$, $m=2$, independent couplings
    $g_j\sim\mathcal N(0,g^2)$ with $g=0.1$, and no
    environment--environment interactions.  Uniform fields have
    $h_j=h_{\rm rms}$ for every witness; for Gaussian fields,
    $h_j\sim\mathcal N(0,h_{\rm rms}^2)$ independently of the couplings.
    Sampling and time averaging follow Fig.~\ref{fig:results_theta_dependence}.
    Shading shows pointwise 95\% confidence intervals obtained by resampling
    independent realizations.}
    \label{fig:supp_theta_fields}
\end{figure}

Local fields change both the axis and frequency of the conditional
rotation, so the no-field invariant $\Lambda$ no longer determines
record quality.  Figure~\ref{fig:supp_theta_fields} isolates this effect
at $p=1/2$, where the no-field response in Fig.~\ref{fig:results_theta_dependence}
is monotonic. A uniform field of rms strength $2g$ produces an interior
maximum instead.
An intermediate angle can therefore provide more information than either
endpoint even after averaging over the signed couplings and time.
Gaussian-distributed fields smooth this maximum, showing that the
angular response also depends on the field distribution.
At $\theta=\pi/2$, all profiles
agree because the field and interaction axes commute.

These angle sweeps complement the nonmonotonic dependence on field strength
and mixedness studied through SBS distance in
Ref.~\cite{MironowiczHorodeckiHorodecki2022SelfEvolution}.
Here we evaluate information in small fragments over a finite window;
the ordering of field-dependent curves need not hold at every time and
does not imply the persistence criterion of
Eq.~\eqref{eq:methods_holevo_min}.
Evaluating four representative parameter points with half the
time spacing changes the plotted mean information by at most $0.004$ bit.

%% file: sections/supplement_mixed_states.tex
\subsection*{Mixed initial witnesses}

\begin{figure}[h!]
    \centering
    \includegraphics[width=.86\textwidth]{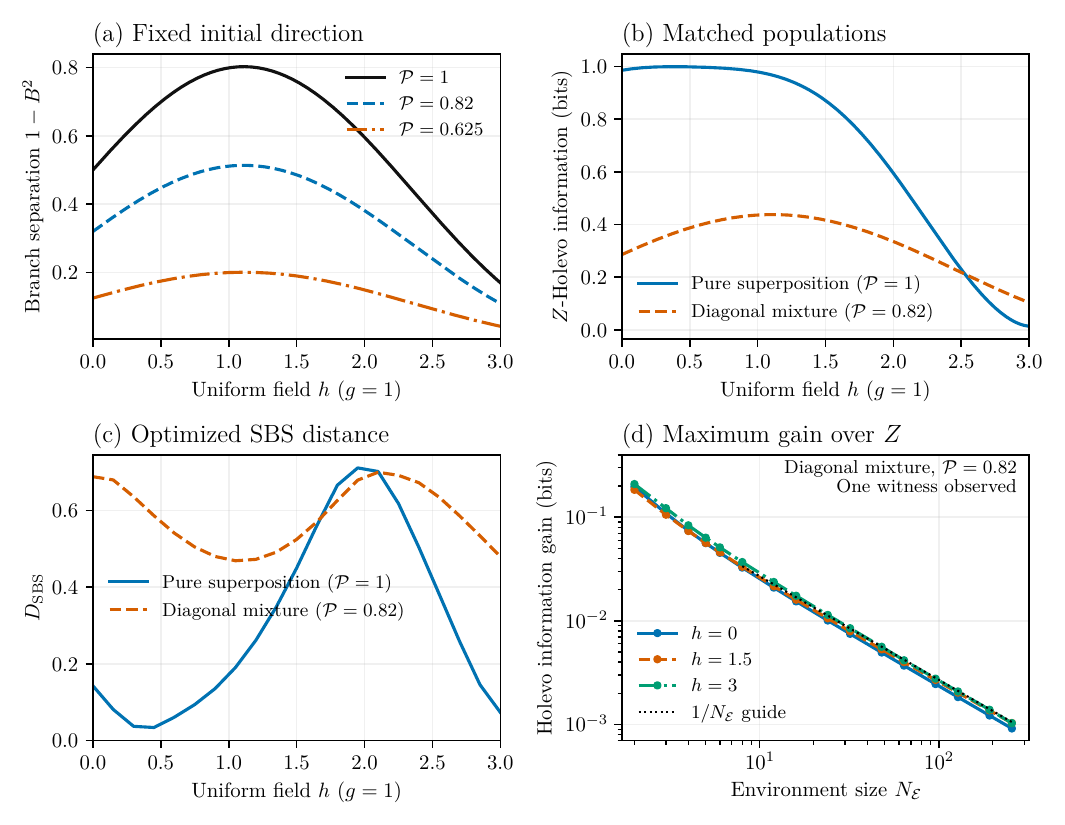}
    \caption{Mixed-state recording and comparison with pure witnesses.
    All cases have unit couplings, a uniform field $h$, no system
    self-Hamiltonian and no interactions between environment qubits.
    The observed fragment is one witness; the other $N_{\Env}-1$ environment
    qubits are traced out. Panels (a--c) use $N_{\Env}=3$ and $t=1$.
    (a) Local branch separation at $\theta=\pi/4$ for diagonal initial states
    with $p_{\rm mix}=0,0.1,0.25$; the $\mathcal P=1$ curve is the pure $|1\rangle$ reference.
    (b,c) The two preparations above at matched populations
    $p=p_{\rm mix}=0.1$ and $\theta=\pi/4$:
    (b) $Z$-Holevo information and (c) numerical minimum SBS distance,
    optimizing both local bases and the mixture weight. The distance is
    the trace norm without a factor of $1/2$, as in
    Ref.~\cite{MironowiczHorodeckiHorodecki2022SelfEvolution}.
    (d) Largest gain over the $Z$ measurement found by optimizing the system
    measurement axis and time over $0\leq t\leq20$, for diagonal witnesses
    with $p_{\rm mix}=0.1$ ($\mathcal P=0.82$) and $\theta=0.9\pi/2$.
    Increasing $N_{\Env}$ adds unobserved witnesses. The dotted line is
    proportional to $1/N_{\Env}$.}
    \label{fig:supp_mixed_comparison}
\end{figure}

We compare diagonal mixed witnesses with the pure preparation of the main
text while keeping their populations equal, $p_{\rm mix}=p$:
\begin{equation}
    \varrho_{\rm mixed}(p)=
    \begin{pmatrix}p&0\\0&1-p\end{pmatrix},
    \qquad
    \varrho_{\rm pure}(p)=
    \begin{pmatrix}p&\sqrt{p(1-p)}\\\sqrt{p(1-p)}&1-p\end{pmatrix}.
\end{equation}
The pure state is an initial superposition; the diagonal mixture removes
its off-diagonal coherence before the system--environment evolution.
Their initial witness purities are $\mathcal P\equiv\operatorname{Tr}(\varrho^2)=1$
and $\mathcal P=p^2+(1-p)^2$, respectively.
Figure~\ref{fig:supp_mixed_comparison}(a) isolates purity reduction at fixed
Bloch-vector direction. Panels (b,c) compare the population-matched states
above, changing both purity and alignment.

Equal branch fidelities need not give equal Holevo information:
at $\theta=h=0$ and $p=p_{\rm mix}=0.25$, both preparations have
$B^2=0.75$, but $\chi_z\simeq0.355$ bits for the pure state and $0.189$ bits
for the mixed state.
Panel (d) shows a finite-environment advantage of rotated measurements,
consistent with $1/N_{\Env}$ scaling over the tested sizes. The dynamically
stable pointer states remain the $Z$ eigenstates; a positive gain does not
establish redundant recording of a rotated observable.

\clearpage
Panels (a,b) and (c) use 301 and 21 field samples, respectively.
SBS searches combine multiple initializations with differential evolution;
global minimality is not certified. Panel (d) searches a hemisphere of
measurement axes on a $9\times64$ polar--azimuthal grid, followed by local
refinement in axis and time. A time grid with spacing $0.1$ is supplemented
near $t=0$ and, for $h=0$, near exact revivals, to resolve peaks that narrow
with increasing size. Nine denser checks change the maxima by less than
$3\times10^{-11}$ bits. Propagators agree with independent full
density-matrix evolution to within $3\times10^{-15}$. The numerical maxima
are not certified global; these single-witness comparisons do not test
persistence or redundancy.

\subsubsection*{Size dependence of the measurement advantage}

The disappearance of the advantage at large environment size also follows
analytically. Write $\Gamma_{\Frag}$ and $\Gamma_{\Rem}$ for the products of
the local decoherence amplitudes $\gamma_j$ over the observed fragment and
its complement, and use $h(x)=H_2((1+x)/2)$ from
Eq.~\eqref{eq:methods_binary_entropy}. For mixed witnesses, these
amplitudes differ from the branch fidelities. After tracing out $\Rem$,
the joint state of $\Ssys$ and $\Frag$ is a controlled unitary applied to
$\rho_{\Ssys}(\Gamma_{\Rem})\otimes\rho_{\Frag}(0)$, where
$\rho_{\Ssys}(c)=\tfrac12\begin{pmatrix}1&c^*\\c&1\end{pmatrix}$.
Unitary invariance of entropy therefore gives
\begin{equation}
    D_z=h(|\Gamma_{\Rem}\Gamma_{\Frag}|)-h(|\Gamma_{\Rem}|).
\end{equation}
Together with Eq.~\eqref{eq:discussion_pointer_gain_bound} and
$1-h(x)\leq x^2$, this bounds the additional information by
\begin{equation}
    0\leq\Delta\chi\equiv\max_{\mathbf n}\chi_{\mathbf n}-\chi_z
    \leq D_z\leq|\Gamma_{\Rem}|^2.
\end{equation}
For identical independent witnesses, let $z=|\gamma(t)|$ and keep the
observed size $m$ fixed. Then $D_z=h(z^{N_{\Env}})-h(z^{N_{\Env}-m})$.
At a fixed time with $z<1$, the preceding bound is exponential in the
number of unobserved witnesses. A bound uniform over time follows by
integrating $dh(z^s)/ds$ and using
$-r\ln r\,\operatorname{atanh}r\leq1/2$ for $0<r<1$:
\begin{equation}
    \sup_t\Delta\chi(t)\leq\frac{m}{2(N_{\Env}-m)\ln2},
    \qquad N_{\Env}>m.
\end{equation}
Thus the advantage vanishes even after time optimization in this
homogeneous model. Exponential decay of that maximum is not implied,
because its location can move toward revivals as the environment grows.

%% file: paper_a_bibliography.bib
@article{Zurek1981PointerBasis,
  author = {Zurek, W. H.},
  title = {Pointer basis of quantum apparatus: Into what mixture does the wave packet collapse?},
  journal = {Physical Review D},
  volume = {24},
  number = {6},
  pages = {1516--1525},
  year = {1981},
  doi = {10.1103/PhysRevD.24.1516},
  url = {https://doi.org/10.1103/PhysRevD.24.1516},
  urldate = {2026-07-17}
}

@article{Zurek2003Decoherence,
  author = {Zurek, Wojciech Hubert},
  title = {Decoherence, einselection, and the quantum origins of the classical},
  journal = {Reviews of Modern Physics},
  volume = {75},
  number = {3},
  pages = {715--775},
  year = {2003},
  doi = {10.1103/RevModPhys.75.715},
  url = {https://doi.org/10.1103/RevModPhys.75.715},
  urldate = {2026-07-17}
}

@article{OllivierPoulinZurek2004ObjectiveProperties,
  author = {Ollivier, Harold and Poulin, David and Zurek, Wojciech H.},
  title = {Objective Properties from Subjective Quantum States: Environment as a Witness},
  journal = {Physical Review Letters},
  volume = {93},
  number = {22},
  pages = {220401},
  year = {2004},
  doi = {10.1103/PhysRevLett.93.220401},
  url = {https://doi.org/10.1103/PhysRevLett.93.220401},
  urldate = {2026-07-17}
}

@article{OllivierPoulinZurek2005EnvironmentWitness,
  author = {Ollivier, Harold and Poulin, David and Zurek, Wojciech H.},
  title = {Environment as a witness: Selective proliferation of information and emergence of objectivity in a quantum universe},
  journal = {Physical Review A},
  volume = {72},
  number = {4},
  pages = {042113},
  year = {2005},
  doi = {10.1103/PhysRevA.72.042113},
  url = {https://doi.org/10.1103/PhysRevA.72.042113},
  urldate = {2026-07-17}
}

@article{BlumeKohoutZurek2006QuantumDarwinism,
  author = {Blume-Kohout, Robin and Zurek, Wojciech H.},
  title = {{Quantum Darwinism}: Entanglement, branches, and the emergent classicality of redundantly stored quantum information},
  journal = {Physical Review A},
  volume = {73},
  number = {6},
  pages = {062310},
  year = {2006},
  doi = {10.1103/PhysRevA.73.062310},
  url = {https://doi.org/10.1103/PhysRevA.73.062310},
  urldate = {2026-07-17}
}

@article{Zurek2009QuantumDarwinism,
  author = {Zurek, Wojciech Hubert},
  title = {{Quantum Darwinism}},
  journal = {Nature Physics},
  volume = {5},
  number = {3},
  pages = {181--188},
  year = {2009},
  doi = {10.1038/nphys1202},
  url = {https://doi.org/10.1038/nphys1202},
  urldate = {2026-07-17}
}

@article{ZwolakQuanZurek2009MixedEnvironment,
  author = {Zwolak, Michael and Quan, H. T. and Zurek, Wojciech H.},
  title = {{Quantum Darwinism} in a Mixed Environment},
  journal = {Physical Review Letters},
  volume = {103},
  number = {11},
  pages = {110402},
  year = {2009},
  doi = {10.1103/PhysRevLett.103.110402},
  url = {https://doi.org/10.1103/PhysRevLett.103.110402},
  urldate = {2026-07-17}
}

@article{ZwolakQuanZurek2010RedundantImprinting,
  author = {Zwolak, Michael and Quan, H. T. and Zurek, Wojciech H.},
  title = {Redundant imprinting of information in nonideal environments: Objective reality via a noisy channel},
  journal = {Physical Review A},
  volume = {81},
  number = {6},
  pages = {062110},
  year = {2010},
  doi = {10.1103/PhysRevA.81.062110},
  url = {https://doi.org/10.1103/PhysRevA.81.062110},
  urldate = {2026-07-17}
}

@article{RiedelZurek2010Everyday,
  author = {Riedel, C. Jess and Zurek, Wojciech H.},
  title = {{Quantum Darwinism} in an Everyday Environment: Huge Redundancy in Scattered Photons},
  journal = {Physical Review Letters},
  volume = {105},
  number = {2},
  pages = {020404},
  year = {2010},
  doi = {10.1103/PhysRevLett.105.020404},
  url = {https://doi.org/10.1103/PhysRevLett.105.020404},
  urldate = {2026-07-17}
}

@article{RiedelZurekZwolak2012RiseFall,
  author = {Riedel, C. Jess and Zurek, Wojciech H. and Zwolak, Michael},
  title = {The rise and fall of redundancy in decoherence and quantum {Darwinism}},
  journal = {New Journal of Physics},
  volume = {14},
  number = {8},
  pages = {083010},
  year = {2012},
  doi = {10.1088/1367-2630/14/8/083010},
  url = {https://doi.org/10.1088/1367-2630/14/8/083010},
  urldate = {2026-07-17}
}

@article{ZwolakRiedelZurek2014Chernoff,
  author = {Zwolak, Michael and Riedel, C. Jess and Zurek, Wojciech H.},
  title = {Amplification, Redundancy, and Quantum {Chernoff} Information},
  journal = {Physical Review Letters},
  volume = {112},
  number = {14},
  pages = {140406},
  year = {2014},
  doi = {10.1103/PhysRevLett.112.140406},
  url = {https://doi.org/10.1103/PhysRevLett.112.140406},
  urldate = {2026-07-17}
}

@article{BrandaoPianiHorodecki2015Generic,
  author = {Brand{\~a}o, Fernando G. S. L. and Piani, Marco and Horodecki, Pawe{\l}},
  title = {Generic emergence of classical features in quantum {Darwinism}},
  journal = {Nature Communications},
  volume = {6},
  number = {1},
  pages = {7908},
  year = {2015},
  doi = {10.1038/ncomms8908},
  url = {https://doi.org/10.1038/ncomms8908},
  urldate = {2026-07-17}
}

@article{KorbiczHorodeckiHorodecki2014Broadcasting,
  author = {Korbicz, J. K. and Horodecki, P. and Horodecki, R.},
  title = {Objectivity in a Noisy Photonic Environment through Quantum State Information Broadcasting},
  journal = {Physical Review Letters},
  volume = {112},
  number = {12},
  pages = {120402},
  year = {2014},
  doi = {10.1103/PhysRevLett.112.120402},
  url = {https://doi.org/10.1103/PhysRevLett.112.120402},
  urldate = {2026-07-17}
}

@article{HorodeckiKorbiczHorodecki2015QuantumOrigins,
  author = {Horodecki, R. and Korbicz, J. K. and Horodecki, P.},
  title = {Quantum origins of objectivity},
  journal = {Physical Review A},
  volume = {91},
  number = {3},
  pages = {032122},
  year = {2015},
  doi = {10.1103/PhysRevA.91.032122},
  url = {https://doi.org/10.1103/PhysRevA.91.032122},
  urldate = {2026-07-17}
}

@article{TuziemskiKorbicz2016BrownianSBS,
  author = {Tuziemski, Jan and Korbicz, Jaros{\l}aw K.},
  title = {Analytical studies of spectrum broadcast structures in quantum {Brownian} motion},
  journal = {Journal of Physics A: Mathematical and Theoretical},
  volume = {49},
  number = {44},
  pages = {445301},
  year = {2016},
  doi = {10.1088/1751-8113/49/44/445301},
  url = {https://doi.org/10.1088/1751-8113/49/44/445301},
  urldate = {2026-07-17}
}

@article{MironowiczKorbiczHorodecki2017Monitoring,
  author = {Mironowicz, P. and Korbicz, J. K. and Horodecki, P.},
  title = {Monitoring of the Process of System Information Broadcasting in Time},
  journal = {Physical Review Letters},
  volume = {118},
  number = {15},
  pages = {150501},
  year = {2017},
  doi = {10.1103/PhysRevLett.118.150501},
  url = {https://doi.org/10.1103/PhysRevLett.118.150501},
  urldate = {2026-07-17}
}

@article{LeOlayaCastro2018Objectivity,
  author = {Le, Thao P. and Olaya-Castro, Alexandra},
  title = {Objectivity (or lack thereof): Comparison between predictions of quantum {Darwinism} and spectrum broadcast structure},
  journal = {Physical Review A},
  volume = {98},
  number = {3},
  pages = {032103},
  year = {2018},
  doi = {10.1103/PhysRevA.98.032103},
  url = {https://doi.org/10.1103/PhysRevA.98.032103},
  urldate = {2026-07-17}
}

@article{LeOlayaCastro2019StrongQD,
  author = {Le, Thao P. and Olaya-Castro, Alexandra},
  title = {Strong Quantum {Darwinism} and Strong Independence are Equivalent to Spectrum Broadcast Structure},
  journal = {Physical Review Letters},
  volume = {122},
  number = {1},
  pages = {010403},
  year = {2019},
  doi = {10.1103/PhysRevLett.122.010403},
  url = {https://doi.org/10.1103/PhysRevLett.122.010403},
  urldate = {2026-07-17}
}

@article{Korbicz2021Roads,
  author = {Korbicz, J. K.},
  title = {Roads to objectivity: Quantum {Darwinism}, Spectrum Broadcast Structures, and Strong quantum {Darwinism}--a review},
  journal = {Quantum},
  volume = {5},
  pages = {571},
  year = {2021},
  doi = {10.22331/q-2021-11-08-571},
  url = {https://doi.org/10.22331/q-2021-11-08-571},
  urldate = {2026-07-17}
}

@article{TouilYanGirolamiEtAl2022Eavesdropping,
  author = {Touil, Akram and Yan, Bin and Girolami, Davide and Deffner, Sebastian and Zurek, Wojciech Hubert},
  title = {Eavesdropping on the Decohering Environment: Quantum {Darwinism}, Amplification, and the Origin of Objective Classical Reality},
  journal = {Physical Review Letters},
  volume = {128},
  number = {1},
  pages = {010401},
  year = {2022},
  doi = {10.1103/PhysRevLett.128.010401},
  url = {https://doi.org/10.1103/PhysRevLett.128.010401},
  urldate = {2026-07-17}
}

@article{MironowiczHorodeckiHorodecki2022SelfEvolution,
  author = {Mironowicz, Piotr and Horodecki, Pawe{\l} and Horodecki, Ryszard},
  title = {Non-Perfect Propagation of Information to a Noisy Environment with Self-Evolution},
  journal = {Entropy},
  volume = {24},
  number = {4},
  pages = {467},
  year = {2022},
  doi = {10.3390/e24040467},
  url = {https://doi.org/10.3390/e24040467},
  urldate = {2026-07-17}
}

@article{DuruisseauTouilDeffner2023PointerStates,
  author = {Duruisseau, Paul and Touil, Akram and Deffner, Sebastian},
  title = {Pointer States and Quantum {Darwinism} with Two-Body Interactions},
  journal = {Entropy},
  volume = {25},
  number = {12},
  pages = {1573},
  year = {2023},
  doi = {10.3390/e25121573},
  url = {https://doi.org/10.3390/e25121573},
  urldate = {2026-07-17}
}

@article{DoucetDeffner2024Classifying,
  author = {Doucet, Emery and Deffner, Sebastian},
  title = {Classifying Two-Body {Hamiltonians} for Quantum {Darwinism}},
  journal = {Physical Review X},
  volume = {14},
  number = {4},
  pages = {041064},
  year = {2024},
  doi = {10.1103/PhysRevX.14.041064},
  url = {https://doi.org/10.1103/PhysRevX.14.041064},
  urldate = {2026-07-17}
}

@article{AcevedoWehrKorbicz2024SBS,
  author = {Acevedo, Alberto and Wehr, Jan and Korbicz, Jaros{\l}aw K.},
  title = {Spectrum broadcast structures from von {Neumann} type interaction {Hamiltonians}},
  journal = {Journal of Mathematical Physics},
  volume = {65},
  number = {12},
  pages = {122102},
  year = {2024},
  doi = {10.1063/5.0208953},
  url = {https://doi.org/10.1063/5.0208953},
  urldate = {2026-07-17}
}

@article{DoucetDeffner2025Compatibility,
  author = {Doucet, Emery and Deffner, Sebastian},
  title = {Compatibility of quantum measurements and the emergence of classical objectivity},
  journal = {Physical Review A},
  volume = {111},
  number = {4},
  pages = {042217},
  year = {2025},
  doi = {10.1103/PhysRevA.111.042217},
  url = {https://doi.org/10.1103/PhysRevA.111.042217},
  urldate = {2026-07-17}
}

@article{ChenZhongLiEtAl2019PhotonicSimulator,
  author = {Chen, Ming-Cheng and Zhong, Han-Sen and Li, Yuan and Wu, Dian and Wang, Xi-Lin and Li, Li and Liu, Nai-Le and Lu, Chao-Yang and Pan, Jian-Wei},
  title = {Emergence of classical objectivity of quantum {Darwinism} in a photonic quantum simulator},
  journal = {Science Bulletin},
  volume = {64},
  number = {9},
  pages = {580--585},
  year = {2019},
  doi = {10.1016/j.scib.2019.03.032},
  url = {https://doi.org/10.1016/j.scib.2019.03.032},
  urldate = {2026-09-03}
}

@article{ZhuSaliceTouilEtAl2025Superconducting,
  author = {Zhu, Zitian and Salice, Kiera and Touil, Akram and Bao, Zehang and Song, Zixuan and Zhang, Pengfei and Li, Hekang and Wang, Zhen and Song, Chao and Guo, Qiujiang and Wang, H. and Mondaini, Rubem},
  title = {Observation of quantum {Darwinism} and the origin of classicality with superconducting circuits},
  journal = {Science Advances},
  volume = {11},
  number = {31},
  pages = {eadx6857},
  year = {2025},
  doi = {10.1126/sciadv.adx6857},
  url = {https://doi.org/10.1126/sciadv.adx6857},
  urldate = {2026-09-03}
}

@article{DoucetDeffner2026NISQ,
  author = {Doucet, Emery and Deffner, Sebastian},
  title = {From compatibility of measurements to exploring quantum {Darwinism} on {NISQ}},
  journal = {Quantum Science and Technology},
  volume = {11},
  number = {3},
  pages = {035008},
  year = {2026},
  doi = {10.1088/2058-9565/ae753f},
  url = {https://doi.org/10.1088/2058-9565/ae753f},
  urldate = {2026-09-03}
}

@article{OllivierZurek2001Discord,
  author = {Ollivier, Harold and Zurek, Wojciech H.},
  title = {Quantum Discord: A Measure of the Quantumness of Correlations},
  journal = {Physical Review Letters},
  volume = {88},
  number = {1},
  pages = {017901},
  year = {2001},
  doi = {10.1103/PhysRevLett.88.017901},
  url = {https://doi.org/10.1103/PhysRevLett.88.017901},
  urldate = {2026-07-17}
}

@article{HendersonVedral2001Correlations,
  author = {Henderson, L. and Vedral, V.},
  title = {Classical, quantum and total correlations},
  journal = {Journal of Physics A: Mathematical and General},
  volume = {34},
  number = {35},
  pages = {6899--6905},
  year = {2001},
  doi = {10.1088/0305-4470/34/35/315},
  url = {https://doi.org/10.1088/0305-4470/34/35/315},
  urldate = {2026-07-17}
}

@article{ZwolakZurek2013DiscordAccessible,
  author = {Zwolak, Michael and Zurek, Wojciech H.},
  title = {Complementarity of quantum discord and classically accessible information},
  journal = {Scientific Reports},
  volume = {3},
  number = {1},
  pages = {1729},
  year = {2013},
  doi = {10.1038/srep01729},
  url = {https://doi.org/10.1038/srep01729},
  urldate = {2026-07-17}
}

@article{Makhlin2002NonlocalProperties,
  author = {Makhlin, Yuriy},
  title = {Nonlocal Properties of Two-Qubit Gates and Mixed States, and the Optimization of Quantum Computations},
  journal = {Quantum Information Processing},
  volume = {1},
  number = {4},
  pages = {243--252},
  year = {2002},
  doi = {10.1023/A:1022144002391},
  url = {https://doi.org/10.1023/A:1022144002391},
  urldate = {2026-07-17}
}

@article{ZhangValaSastryWhaley2003Geometric,
  author = {Zhang, Jun and Vala, Jiri and Sastry, Shankar and Whaley, K. Birgitta},
  title = {Geometric theory of nonlocal two-qubit operations},
  journal = {Physical Review A},
  volume = {67},
  number = {4},
  pages = {042313},
  year = {2003},
  doi = {10.1103/PhysRevA.67.042313},
  url = {https://doi.org/10.1103/PhysRevA.67.042313},
  urldate = {2026-07-17}
}

@misc{Lasek2026pySLQubits,
  author = {Lasek, Aleksander},
  title = {{pySL\_qubits}: Quantum {Darwinism} simulation and analysis software},
  year = {2026},
  howpublished = {GitHub, version 1.2.0},
  url = {https://github.com/ALasek/pySL_qubits-public/releases/tag/paper-a-v1.2.0},
  note = {Git tag paper-a-v1.2.0}
}

@article{FellerRousselFrerotDegiovanni2021Comment,
  author = {Feller, Alexandre and Roussel, Benjamin and Fr{\'e}rot, Ir{\'e}n{\'e}e and Degiovanni, Pascal},
  title = {Comment on ``{Strong Quantum Darwinism and Strong Independence Are Equivalent to Spectrum Broadcast Structure}''},
  journal = {Physical Review Letters},
  volume = {126},
  number = {18},
  pages = {188901},
  year = {2021},
  doi = {10.1103/PhysRevLett.126.188901},
  url = {https://doi.org/10.1103/PhysRevLett.126.188901},
  urldate = {2026-09-14}
}

@book{Helstrom1976QuantumDetection,
  author = {Helstrom, Carl W.},
  title = {Quantum Detection and Estimation Theory},
  publisher = {Academic Press},
  address = {New York},
  year = {1976},
  isbn = {978-0-12-340050-5},
  url = {https://shop.elsevier.com/books/quantum-detection-and-estimation-theory/helstrom/978-0-12-340050-5},
  urldate = {2026-09-07}
}

@incollection{Levitin1995OptimalMeasurements,
  author = {Levitin, L. B.},
  title = {Optimal Quantum Measurements for Two Pure and Mixed States},
  booktitle = {Quantum Communications and Measurement},
  editor = {Belavkin, V. P. and Hirota, O. and Hudson, R. L.},
  publisher = {Springer},
  address = {Boston, MA},
  pages = {439--448},
  year = {1995},
  doi = {10.1007/978-1-4899-1391-3_43},
  url = {https://doi.org/10.1007/978-1-4899-1391-3_43},
  urldate = {2026-09-07}
}

@article{ZwolakRiedelZurek2016SpinEnvironments,
  author = {Zwolak, Michael and Riedel, C. Jess and Zurek, Wojciech H.},
  title = {Amplification, Decoherence and the Acquisition of Information by Spin Environments},
  journal = {Scientific Reports},
  volume = {6},
  number = {1},
  pages = {25277},
  year = {2016},
  doi = {10.1038/srep25277},
  url = {https://doi.org/10.1038/srep25277},
  urldate = {2026-09-14}
}
